\documentclass[reprint,
nofootinbib,
 amsmath,amssymb,
 aps
floatfix,
]{revtex4-2}

\usepackage{graphicx}
\usepackage{dcolumn}
\usepackage{bm}
\usepackage{physics}
\usepackage{color}
\usepackage{graphicx}
\usepackage{booktabs}

\usepackage{float} 

\usepackage{hyperref}
\usepackage{nameref}
\usepackage{cleveref}
\usepackage[bottom]{footmisc}
\usepackage[mathlines]{lineno}

\begin{document}

\preprint{APS/123-QED}

\title{Probing inflationary particle production with the CMB power spectrum} 
\author{Hidde T.~Jense}
\email{jenseh@cardiff.ac.uk}
\affiliation{School of Physics and Astronomy, Cardiff University, The Parade, CF24 3AA Cardiff, Wales, UK}

\author{Luca H.~Abu El-Haj}
\email{lha2126@columbia.edu}
\affiliation{Department of Physics, Columbia University, New York, NY 10027, USA}

\author{J.~Colin Hill}
\affiliation{Department of Physics, Columbia University, New York, NY 10027, USA}

\author{Oliver~H.\,E.~Philcox}

\affiliation{Leinweber Institute for Theoretical Physics at Stanford, 382 Via Pueblo, Stanford, CA 94305, USA}
\affiliation{Kavli Institute for Particle Astrophysics and Cosmology, 382 Via Pueblo, Stanford, CA 94305, USA}

\begin{abstract}
\noindent Particle production is common to many microphysical models of inflation and can imprint observable features in the cosmic microwave background (CMB) anisotropies. We consider a scenario in which the inflaton couples to an extremely massive field ($m_\chi \gtrsim \mathcal{O}(100 H_I)$, where $H_I$ is the inflationary Hubble scale). In this model, particle production happens in a burst at a characteristic conformal time, $\eta_*$, which sources localized features in the CMB. In this paper, we compute the full temperature and polarization two-point functions for this model.  We then search for these features in CMB power spectrum data from \emph{Planck} and the Atacama Cosmology Telescope (ACT), with the latter allowing access to features on smaller angular scales. In the joint analysis of \emph{Planck} and ACT data, we find a mild $\sim 2 \sigma$ hint for a signal induced by this inflationary model on scales $3 \,\, \text{Mpc}\leq\eta_*\leq 10 \,\, \text{Mpc}$, though this hint is not present at a statistically significant level in either dataset when analyzed individually. Using a Fisher forecast, we find that these features should be observable at the $3-5\sigma$ level for a Simons Observatory-like experiment, if they are indeed real. We also compare our power-spectrum-based constraints to previous matched-filter-based bounds on this model \cite{El-Haj:2025zbe}.  For sufficiently light particles ($m_\chi \lesssim 200 H_I$), the power spectrum yields tighter constraints by more than an order of magnitude, but in the higher-mass regime where particle production is rare, the matched-filter approach provides stronger bounds.
\end{abstract}

\maketitle

\section{\label{sec:level1}Introduction}

\noindent Observations of the cosmic microwave background (CMB) provide tight constraints on the power spectrum of primordial scalar fluctuations. Anisotropies in the distribution of photon temperatures and polarizations were seeded by inflation, and as such, the CMB provides a window across several orders of magnitude in scale to probe the primordial power spectrum. These observations suggest that, consistent with the simplest inflationary models, inflation generated Gaussian and almost scale-invariant initial conditions. Among the central goals of modern cosmology is to understand the underlying physics of inflation, and the detection of any deviations from Gaussianity or a non-power-law primordial power spectrum would provide extremely useful clues in this quest. 

Given that we are far from isolating the fundamental physics of inflation, a natural way to progress is to identify features common to a variety of well-motivated particle physics theories, and explore their phenomenological features in the CMB and other cosmological observables. One such feature is the non-adiabatic production of particles during the inflationary epoch. This has garnered a significant theoretical and phenomenological effort to constrain the production of particles with masses near the inflationary Hubble scale, $m_\chi \sim H_I$, in the CMB and large-scale structure (LSS), known as the ``cosmological collider physics'' program, where particle production characteristically appears as an oscillatory feature in the squeezed limit of low-order correlators~\citep[e.g.,][]{arkanihamed2015cosmologicalcolliderphysics, philcox2025searchinginflationaryphysicscmb,Sohn_2024,Chen_2010,Lee:2016vti,Suman:2025tpv,Philcox:2026njr,Kumar:2026dih}. 

The study of extremely massive particles ($m_\chi\gg H_I$), though well motivated from a particle physics standpoint, has received comparatively less attention, but offers novel phenomenology, with the features associated with these fields depending heavily on the masses and the production mechanism. The thermal production of massive particles is Boltzmann-suppressed with $m_\chi/H_I$, and as such, for extremely massive fields, particle production is rare. That said, there can be moments and production mechanisms for which this particle production is enhanced. In the following, we consider non-adiabatic particle production where, instead of being suppressed by $m_\chi/H_I$ the production rate is suppressed by $m_\chi/\dot{\phi}^{\frac{1}{2}}$, making particle production favorable in some regions of parameter space. It has been shown that sufficiently massive particles, upon production, induce localized position-space features in the CMB and in the late-time matter density field~\cite{El-Haj:2025zbe,kumar2025earlygalaxiesrareinflationary,philcox2025searchinginflationaryphysicscmb,Kim:2023wuk,Kim:2021ida,Flauger_2017,M_nchmeyer_2019}.

In this paper, we consider a theory in which the effective mass of a heavier state passes briefly through a minimum, leading to a burst of particle production at a single point in time. One can then look for evidence of this production in late-time cosmological observables. Various approaches have been employed to search for such features, the most successful being the adoption of a matched-filter profile-finding statistic, whose output is then interpreted using a Poissonian likelihood~\cite{Philcox:2024jpd,El-Haj:2025zbe} (see also \citep{Kim:2021ida,Kim:2023wuk}). That said, in a regime where particle production is less heavily suppressed (for example, $m_\chi\lesssim200 H_I$, for which we expect $\sim100$ events for natural couplings~\cite{El-Haj:2025zbe}), one might expect a more conventional power spectrum search to provide stronger constraints on the theory. This is of course just a simple statement of the central limit theorem: when the signal manifests itself as rare events, the one-point function dominates the constraints, whereas if the signal is common the power spectrum dominates.  

The CMB power spectrum directly probes the primordial power spectrum and thus constrains the physics of inflation. Much of this work has been performed using the \emph{WMAP} and \emph{Planck} satellite mission data~\cite[e.g.,][]{WMAP:2003syu,Planck2018Cosmology,Planck2018Inflation}; more recently, ground-based experiments such as the Atacama Cosmology Telescope (ACT)~\cite{ACT:2025fju} and South Pole Telescope (SPT)~\cite{SPT-3G:2025bzu} have probed the CMB temperature and polarization anisotropies at higher resolution, casting light on smaller-scale primordial features. Additional probes, such as the CMB lensing power spectrum, measured by \emph{Planck}~\cite{Carron_2022}, ACT~\cite{Qu_2024,Madhavacheril_2024}, and SPT~\cite{SPT-3G:2025bzu}, or baryon acoustic oscillation (BAO) distance measurements from DESI~\cite{DESI:2025zpo,DESI:2025zgx} complement these datasets by breaking parameter degeneracies, even if they are less sensitive (or insensitive in the case of BAO) to the primordial power spectrum themselves.  In addition, one can directly constrain primordial features via their signatures in the matter power spectrum via ``full-shape'' analyses of the galaxy power spectrum~\cite[e.g.,][]{Beutler:2019ojk,Mergulhao:2023ukp,martinezsomonte2026primordialpowerspectrumreconstructions,green2026extendingcosmologicalcollidernew}.  Combinations of these datasets allow for analyses that optimize both the range of scales probed for inflationary features and constraints on the background evolution \citep[e.g.,][]{ACT:2025tim}.  In the following, we undertake the first power-spectrum-based search for features induced by the inflationary particle production mechanism detailed in~\cite{Flauger_2017,Kim:2021ida}. Complementary work, in~\citep{Nervalinprep}, investigates a similar search for features from a broad class of oscillatory templates in a dataset similar to the one investigated here. A subset of these templates is derived from single and multiple bursts of massless particle production during inflation. The work in this paper is focused on the phenomenology of the massive-particle-production model from~\citep{Kim:2021ida, philcox2025searchinginflationaryphysicscmb,El-Haj:2025zbe} and building on work from there, whereas the work in~\citep{Nervalinprep} studies a different class of models, searching for a more generalized oscillatory form in the primordial power spectrum. The two works complement each other to provide a wide overview of observational constraints on non-standard inflationary models.

The remainder of this work is organized as follows. In \autoref{sec:theory}, we discuss the modification to the power spectrum induced by particle production from both a primordial and late-time approach. In \autoref{sec:data}, we describe the data sets used to search for the particle production signature, before outlining our procedure for this search and how we derive bounds on the associated inflationary physics in \autoref{sec:sims}. In \autoref{sec:Results}, we describe the results of our search and compare them to previous matched-filter-based approaches~\cite{Philcox:2024jpd,El-Haj:2025zbe} and to prospective bounds derived from future experiments. We conclude in \autoref{sec:summary}. Throughout, we use the $(-,+,+,+)$ metric signature convention and adopt the following Fourier conventions:
\begin{align}
    f(\mathbf{x})&=\int \frac{\dd^3 k}{(2\pi)^3}f(\mathbf{k})e^{-i\mathbf{k}\cdot\mathbf{x}}\equiv\int_{\mathbf{k}}f(\mathbf{k})e^{-i\mathbf{k}\cdot\mathbf{x}}\\
    f(\mathbf{k})&=\int \dd^3 xf(\mathbf{x})e^{i\mathbf{k}\cdot\mathbf{x}}\equiv \int_{\mathbf{x}}f(\mathbf{x})e^{i\mathbf{k}\cdot\mathbf{x}} \nonumber
\end{align}
Similarly, we define an appropriately normalized Dirac delta-function, $\delta_{\mathbf{k}}\equiv(2\pi)^3\delta^{D}(\mathbf{k}),~\delta_{\mathbf{x}}\equiv\delta^{D}(\mathbf{x})$, such that $\int_{\mathbf{x}}\delta_{\mathbf{x}}=\int_{\mathbf{k}}\delta_{\mathbf{k}}=1$.

\section{Theoretical Background}
\label{sec:theory}

\noindent Particle production can happen in a plethora of multifield inflationary scenarios. Here we restrict to a scenario in which heavy particles $\chi$ are produced in a narrow window around the minimum of the time-dependent effective mass of the field, $m_\chi(\phi)$~\cite{Kim:2021ida}. If the effective mass reaches its minimum at some inflaton field value $\phi_*\equiv\phi(t_*)$, we can Taylor-expand the mass,
\begin{align}
    m_\chi(\phi)\simeq m_{\chi}(\phi_*)+\frac{1}{2}\dv[2]{m_\chi}{\phi} \biggr\rvert_{\phi_*} (\phi-\phi_*)^2~. 
\end{align}
We can then realize this as an interaction term in the Lagrangian via 
\begin{align}
    \mathcal{L}\supset -\frac{1}{2} \left(\partial\chi \right)^2-\frac{1}{2} \left(m_0^2+(g\phi-\mu)^2\right)\chi^2 \,,
    \label{eqn:lagrangian}
\end{align}
where we have defined the coupling constants $g^2\equiv m_0\dv[2]{m_\chi}{\phi}\rvert_{\phi_*}$ and
$\mu\equiv g\phi_*$, as well as the minimum mass $m_0\equiv m_\chi(\phi_*)$. Given the Lagrangian, we may compute the number density of produced particles $n$ using standard Bogoliubov techniques (see, e.g.,~\cite{Kim:2021ida,Flauger_2017} for a pedagogical review),
\begin{equation}
    n=\left( \frac{g\dot{\phi}_0}{4\pi^2} \right)^{\frac{3}{2}}\exp{-\frac{\pi(m_0^2-2H_I^2)}{g|\dot{\phi}_0|}}\,,
    \label{eqn:number}
\end{equation}
with $\dot{\phi}_0$ the slow rolling velocity. Working in the regime where $m_0\gg H_I$, these particles almost immediately become non-relativistic after production and induce local curvature perturbations, as computed in~\cite{Kim:2021ida}. This can be thought of as the interactions between $\phi$ and $\chi$ locally slowing down the inflaton field's evolution, which in turn leads to hot and cold spots\footnote{For concision, we generally refer to these as simply ``hotspots'', though they can be hot or cold.} in the late-time CMB, depending on the the microphysical parameters. Naturally, these hotspots imprint a new scale on the CMB (that of the conformal time at particle production). The remainder of this work is devoted to computing the effects of these hotspots on the CMB and searching for them in observational data. 

\subsection{Backreaction Constraints}
\noindent We first investigate the regime of validity of our theory, following \cite{Kim:2021ida}. In particular, in the regime of interest for a power spectrum search, we expect a large number of particles to be produced, and as such an important constraint is obtained by enforcing that the particle production mechanism does not appreciably modify the background evolution of the Universe during inflation, which we assume to be (quasi-)de Sitter. Our inflaton background evolution follows the Klein-Gordon equation:
\begin{align}
   \ddot{\phi}+3H_I\dot{\phi}+\pdv{V}{\phi}=0\,.
\end{align}
Thus if we want the dominant contribution to the background evolution to be unchanged compared to the non-coupled case, we require $\pdv{V}{\phi}\simeq \pdv{V_\phi}{\phi}$, where $V_\phi$ represents the part of the potential containing only inflaton self-interactions. Noting that
\begin{align}
    \pdv{V}{\phi}=\pdv{V_\phi}{\phi}+g(g\phi-\mu)\ev{\chi^2}\,,
\end{align}
this requirement becomes:
\begin{align}
    \pdv{V_\phi}{\phi}\gg g(g\phi-\mu)\ev{\chi^2} \,.
\end{align}
Then, using the fact that $3 H_I\dot{\phi}_0\simeq \pdv{V_\phi}{\phi}$ by the slow-roll condition, and denoting the background inflaton field evolution $\phi_0$, we can write
\begin{align}
    H_I\dot\phi_0\gg gm_0\ev{\chi^2},
\end{align}
additionally recalling that $g\phi-\mu\sim m_0$ near the minimum of the effective mass of $\chi$. Finally, we insert the expected number density of produced $\chi$ particles, $n= \int_\mathbf{k}\abs{\beta}^2\sim m_0 \ev{\chi^2}$, where $\beta$ is the $\beta-$Bogoliubov coefficient, to find
\begin{align}
    \frac{\dot{\phi}_0}{H_I^2}\simeq 3600\gg \frac{g n}{H_I^3}\,.
\end{align}
Here we have inserted the approximate numerical value for $\frac{\dot{\phi}_0}{H_I^2}$ obtained from the large-scale CMB fluctuation amplitude. This defines an explicit, if approximate, parameter regime of interest for our power spectrum search. As we will see in \autoref{sec:Results}, for perturbative values of $g$ our constraints lie significantly below these limits, implying that any induced backreaction is negligible. 

\subsection{Computing the Power Spectrum Contribution}
\noindent In the following, we derive the hotspot-induced changes to the CMB power spectrum. We consider both the  contribution from a single hotspot and a correlated pair, noting that conservation of momentum in a homogeneous FRW background requires that the particles be produced in pairs. We do this in two ways. First, we compute the modified two-point function at the primordial level before before mapping this to the CMB. Secondly, we compute the two-point function as a late-time observable, by analogy to the halo model, which yields consistent results.

\subsubsection{Primordial Approach}
\noindent We start by deriving the contribution of massive particles to the primordial power spectrum. The production of particles is treated as Poissonian and is, at least to leading order, uncorrelated with inflaton fluctuations. As noted above, conservation of momentum implies that each production event creates two massive particles; thus, we obtain a Poisson distribution of events, each of which produces two hotspots.  We derive the pairwise contribution to the power spectrum using the ``in-in'' formalism, as described in Appendix~\ref{Appendix:appendix A}. Our formalism bears similarity to the astrophysical ``halo model'' \cite[e.g.,][]{Cooray:2002dia,Hill_2013}, involving ``one-'' and ``two-halo'' contributions sourced, respectively, by the uncorrelated and correlated components of the pairwise events.

To begin, we require the action of two massive particles produced during inflation, which is a simple extension of the one-particle action presented in \cite{Kim:2021ida}:
\begin{align}
\label{eq:2paction}
S_{\text{2-particle}}=\int_{\eta_*}^0 \dd \eta\int_{\mathbf{k}} \partial_\eta\zeta\frac{m_\chi(\eta)e^{i\vb{k}\cdot \vb{x}_{\rm HS}}}{H_I}[1+e^{i\vb{k}\cdot \vb{d}}]\,,
\end{align}
where $\eta_*$ is the conformal time of particle production, $\zeta(\mathbf{x},\eta)$ is the curvature perturbation, $\mathbf{x}_{\rm{HS}}$ is the position of the first particle, and $\vb{d}$ gives the separation between the two hotspots. The value of $|\vb{d}|$ depends heavily on the microphysics of the problem, but to enforce causality we require $\abs{\vb{d}}\leq\abs{\eta_*}$\cite{Kim:2021ida}\footnote{Herein we adopt notation similar to that of \cite{Kim:2021ida} where $\eta_*$ is the absolute value of the conformal time during particle production.}. This induces the connected two-point function, 
\begin{widetext}
\begin{align}
\label{eq:powerspec}
    \langle \zeta_{\mathbf{k}_1}\zeta_{\mathbf{k}_2}\rangle_{\text{PP}} &=\frac{g^2H_I^4}{\dot{\phi}_0^2k_1^3k_2^3}\big(\text{Si}(k_1\eta_*)-\sin(k_1\eta_*)\big)\big(\text{Si}(k_2\eta_*)-\sin(k_2\eta_*)\big) e^{-i(\mathbf{k_1}+\mathbf{k_2})\cdot\mathbf{x}_{\text{HS}}}\nonumber \\&\times
    [1+e^{-i(\mathbf{k_1}+\mathbf{k_2})\cdot\mathbf{d}}+e^{-i\mathbf{k}_1\cdot\mathbf{d}}+e^{-i\mathbf{k}_2\cdot\mathbf{d}}]\,,
\end{align}
\end{widetext}
derived in Appendix~\ref{Appendix:appendix A}, where Si is the sine integral.  Note that the first two contributions in the second line, which define the ``one-hotspot'' term, agree with the result of \cite{kumar2025earlygalaxiesrareinflationary}. Eq.~\eqref{eq:powerspec} describes the effect of one massive particle production event; the full spectrum can be computed by averaging over position through the volume integral~$1/\mathcal{V}\int \dd ^3\mathbf{x}_{\text{HS}}$ and weighting by the expected number density of particle production. This yields
\begin{align}
    \langle \zeta_{\mathbf{k}_1}\zeta_{\mathbf{k}_2}\rangle_{\text{PP}} &=\delta_{\vb{k_1}+\vb{k_{2}}}\frac{g^2H_I^4 n_{\text{co}}}{\dot{\phi}^2k_1^6}(\text{Si}(k_1\eta_*)-\sin(k_1\eta_*))^2\nonumber \\&\times[1+\cos{\mathbf{k}_1\cdot \mathbf{d}}] \,, 
\end{align} 
where $n_{\text{co}}$ is the comoving number density of produced particles: $n_{\rm co} = n/a^3 = n/(H_I\eta_*)^3$. In analogy with the halo model, the cosine term plays the role of the ``two-halo'' term (depending explicitly on the particle separation), while the other contribution corresponds to the ``one-halo'' term (depending only on the total number of particles). We refer to these as one- and two-hotspot terms for the remainder of this work, so as to avoid explicit confusion with the halo model. 

Finally, we can then average over the hotspot separation $\vb{d}$. The angular average is simply given by $\frac{1}{2}\int_{-1}^1 \cos(k_1\abs{\mathbf{d}}\mu) \dd \mu=\frac{\sin(k_1\abs{\mathbf{d}})}{k_1\abs{\mathbf{d}}}$. Avoiding a detailed treatment of microphysics, we assume $\abs{\mathbf{d}}$ to be uniformly distributed in the range $[0,\abs{\eta_*}]$. 
which leads to the primordial two-point function
\begin{align}
   \langle \zeta_{\mathbf{k}_1}\zeta_{\mathbf{k}_2}\rangle_{\text{PP}}&=\delta_{\mathbf{k}_1+\mathbf{k}_2}\frac{g^2H_I^4 n_{\text{co}}}{\dot{\phi}^2_0k_1^6} \left( \text{Si}(k_1\eta_*)-\sin(k_1\eta_*) \right)^2 \nonumber \\&\times\left[ 1 + 3\frac{\sin(k_1 \eta_*) - k \eta_* \cos(k_1 \eta_*)}{(k_1\eta_*)^3} \right]\,,
\end{align}
where the second term in the brackets describes the two-hotspot contribution. As in \cite{kumar2025earlygalaxiesrareinflationary}, we can simply write this as a multiplicative correction to the primordial power spectrum,
\begin{widetext}
    \begin{align}
    P_{\Lambda\text{CDM}+\text{PP}}(k)=P_{\text{std}}(k) \left(1+2 g^2 P_* \frac{(\text{Si}(k\eta_*)-\sin(k\eta_*))^2}{(k\eta_*)^3} \left[1+3\frac{\sin(k\eta_*)-k\eta_*\cos(k\eta_*)}{(k\eta_*)^3} \right] \right) \,, \label{eq:pps-full-expression}
\end{align}
\end{widetext}
where we have defined\footnote{While we assume scale invariance, for simplicity, throughout the primordial hotspot computation, for the purposes of our analysis we use the standard $P_{\rm{std}}\propto k^{n_s-1}$ expression.}, 
\begin{align}
    &P_{\text{std}}(k)=\frac{H_I^4}{2\dot{\phi}_0^2k^3} \\
    &P_*=\frac{n}{H_I^3}=\left(\frac{g\dot{\phi}_0}{4\pi^2 H_I^2}\right)^{3/2}e^{-\pi(m_0^2-2H_I^2)/(g\abs{\dot{\phi_0}})}\,.
    \label{eq:pps-amplitude}
\end{align}

Finally, the CMB power spectrum takes the form
\begin{align}
    C_\ell^{XY}=\frac{2}{\pi}\int &\dd k k^2\mathcal{T}_\ell^X(k)\mathcal{T}^Y_\ell(k) P_{\Lambda\rm{CDM}+\rm{PP}}(k) \, ,
    \label{eq:pps-CMBLOS}
\end{align}
where $\mathcal{T}_\ell^X(k)$ denotes the CMB transfer function for $X$-modes, with $X\in\{T,E\}$, which can be computed with a standard Einstein-Boltzmann solver.
Above, $P_{\rm{\Lambda CDM}+\rm{PP}}(k)$ denotes the full contribution to the power spectrum from both the standard power-law $\Lambda$CDM primordial spectrum and the particle production hotspots.

The scale dependence of the hotspot feature induces a peak in the primordial power spectrum at $k_* = \pi / \eta_*$, with a tapered shape due to the $(k \eta_*)^3$ fall-off at smaller and larger scales. The two-hotspot term modifies the amplitude of the feature on scales below $k_*$, where the size of this modification depends on the relevant $\eta_*$ value.  

\begin{figure}[t]
    \centering
    \includegraphics[width=\linewidth]{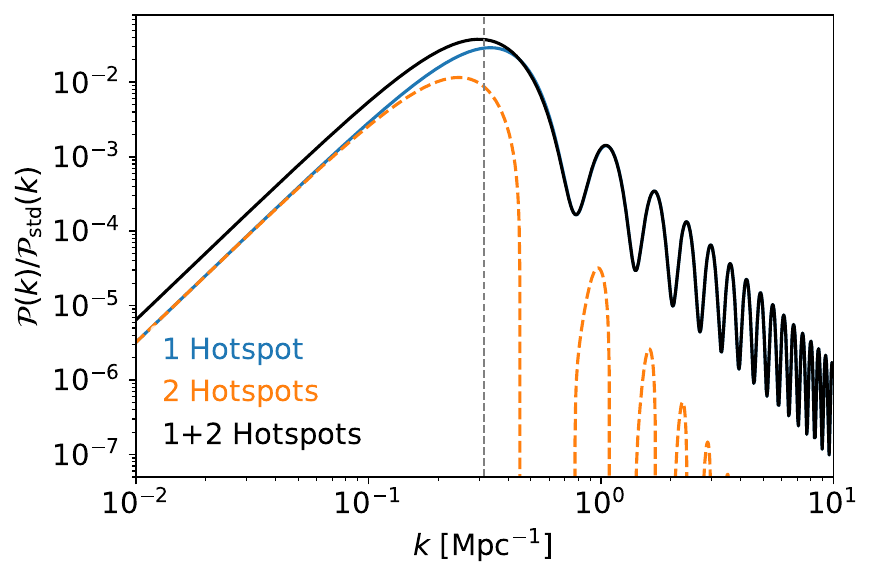}
    \caption{
    The induced contribution of a hotspot feature to the primordial power spectrum. We show the fractional contribution in the modified power spectrum including only the 1-hotspot term (\textbf{blue}), the 2-hotspot term (\textbf{orange dashed}), and both the 1- and 2-hotspot terms (\textbf{black}). The parameters chosen are $\eta_* = 10 \, \rm{Mpc}$ and $2 g^2 P_* = 0.26$ (chosen from the \texttt{P-ACT} constraints found in \autoref{tab:pps-upper-limits-values}) The vertical dashed line shows $k_* = \pi / \eta_*$, which is the wavenumber where the contribution is approximately maximal. The changes in observables arising from this contribution are shown in~\autoref{fig:pps-fullcamb} and~\autoref{fig:matter-power}.}
    \label{fig:pps-modified}
\end{figure}

\begin{figure*}
    \centering
    \includegraphics[width=\linewidth]{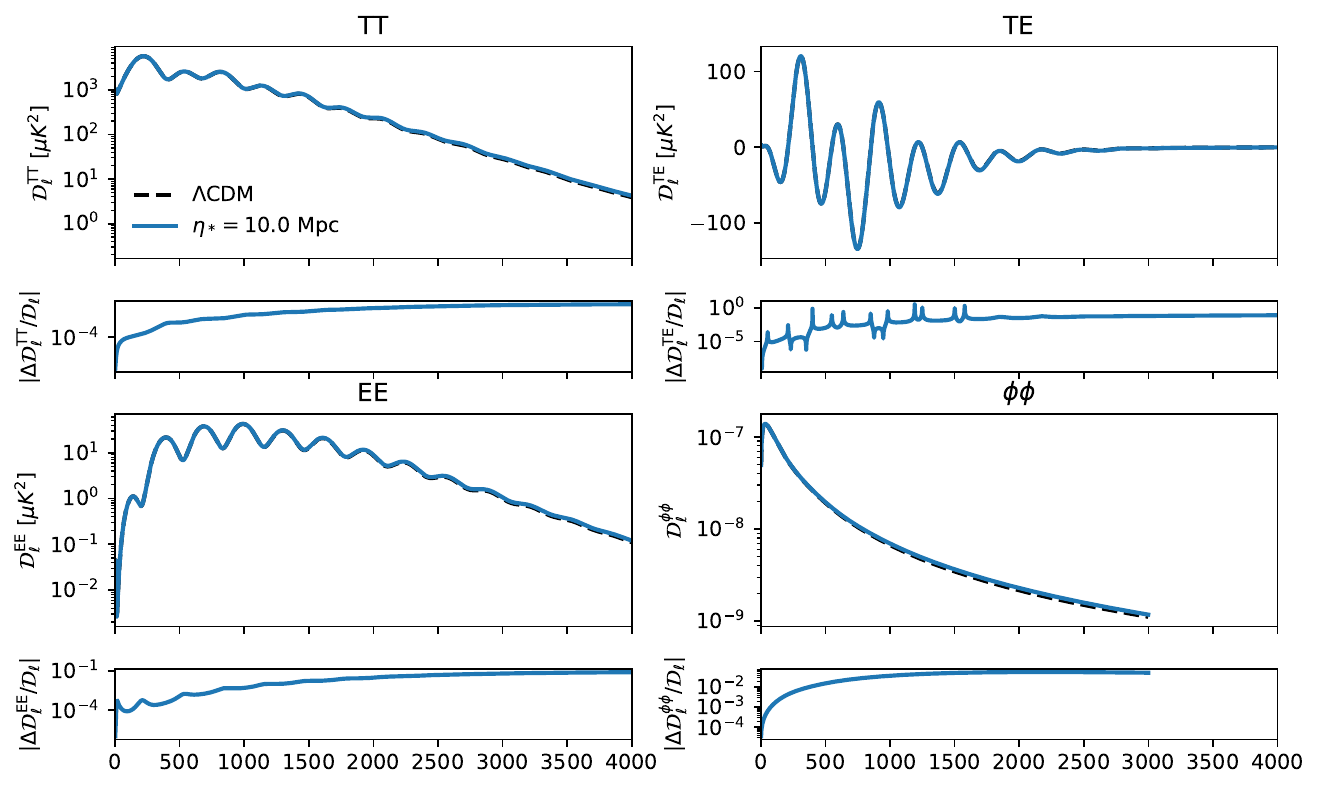}
    \caption{Changes in the CMB power spectra arising from the primordial power spectrum contribution shown in~\autoref{fig:pps-modified}. We show results for the temperature (TT, \textbf{top-left}), polarization (EE, \textbf{bottom-left}), cross (TE, \textbf{top-right}), and lensing potential ($\phi\phi$, \textbf{bottom-right}). In each case, we show a comparison of the $\Lambda$CDM prediction (\textbf{black dashed} lines) to that obtained when including primordial hotspot contributions (\textbf{blue solid} lines). The larger panels show a comparison between the spectra, whereas the smaller panels show the absolute value of the fractional difference relative to $\Lambda$CDM. The cosmological parameters are based on the \texttt{P-ACT} best-fit~\cite{ACT:2025tim}: $\{\theta_{\rm MC} = 104.0754 \times 10^{-4}, \omega_\text{b}=2.25 \times 10^{-2}, \omega_{\text{cdm}}=11.919 \times 10^{-2}, \tau_{\text{reio}}=6.01 \times 10^{-2}, \sum m_\nu=0.06~\text{eV}, A_s=2.123\times 10^{-9}, n_s=0.9712\}$, while the parameters for the hotspot contribution are $\eta_* = 10 \, \rm{Mpc}$ and $2 g^2 P_* = 0.26$, the same as in~\autoref{fig:pps-modified}.
    }
    \label{fig:pps-fullcamb}
\end{figure*}

\begin{figure}
    \centering
    \includegraphics[width=\linewidth]{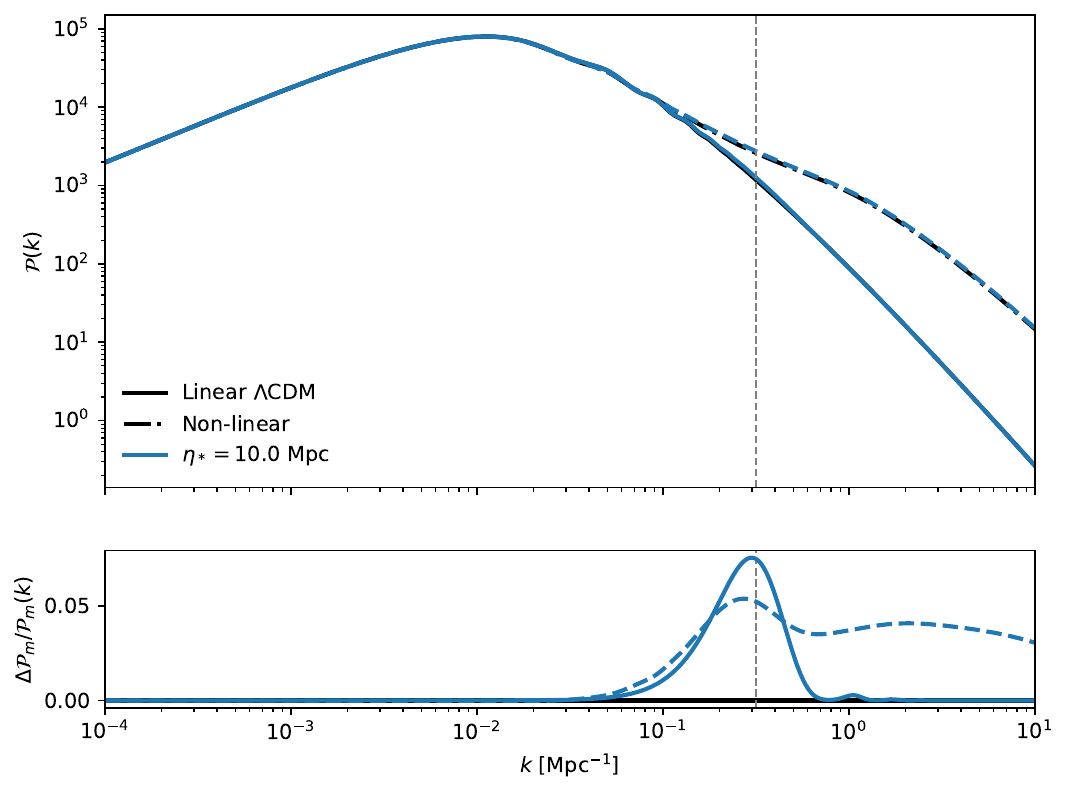}
    \caption{Changes in the linear and non-linear matter power spectra coming from the primordial contribution shown in~\autoref{fig:pps-modified}. We show the linear (\textbf{solid} lines) and non-linear (\textbf{dashed} lines) matter power spectra for $\Lambda$CDM (\textbf{black} lines) and when including the primordial hotspot contribution (\textbf{blue} lines). The larger panel shows a comparison of the observables, whereas the lower panel shows the fractional difference relative to $\Lambda$CDM. The vertical dashed line indicates $k_* = \pi / \eta_*$, which is the wavenumber sourcing the largest hotspot contribution to the primordial power spectrum. The cosmological and hotspot parameters are identical to those used in~\autoref{fig:pps-fullcamb}. Note that the non-linear model simply uses the \texttt{mead2020} preset from \texttt{camb}, which may not be highly accurate for this model as it was calibrated for $\Lambda$CDM cosmologies.  However, the observables studied in this work are dominated by linear-theory (or background-only) contributions.
    }
    \label{fig:matter-power}
\end{figure}

\vspace{6pt}

\par We show the effects of our model in~\Cref{fig:pps-modified,fig:pps-fullcamb,fig:matter-power}. \autoref{fig:pps-modified} shows the contribution to the primordial power spectrum from a feature of interest. We differentiate between the main 1 Hotspot and the additional 2 Hotspot terms. For a hotspot feature with a given $\eta_*$, we note that the contribution to the primordial power spectrum maximizes around $k = \pi / \eta_*$. In \autoref{fig:pps-fullcamb} we then show how this feature affects the CMB power spectrum in temperature, E-mode polarization, and lensing. We show both a comparison between this observable and $\Lambda$CDM, and the fractional difference between the two models. We note that the fractional difference is high for the EE and lensing power spectra, showing the importance of these datasets from ACT DR6 when adding them to the \emph{Planck} temperature power spectra. Finally, in \autoref{fig:matter-power}, we show the effects on the linear and non-linear matter power spectra. We note that the contribution to the linear matter power spectrum is directly proportional to the feature itself, while the non-linear matter power spectrum smears out the feature to smaller scales. Whilst the non-linear modeling used by cosmological theories such as \texttt{camb} are calibrated around $\Lambda$CDM cosmologies and may thus not accurately represent the non-linear feedback at these scales, we do not directly probe the matter power spectrum in this work, and the observables we do use are dominated by linear contributions.

\subsubsection{Late-Time Approach}
\noindent Next, we show that we can obtain the same result starting from the late-time temperature field and adding spatially local (pairwise) hotspots. This approach connects cleanly to the position-space searches considered in previous works \citep[e.g.,][]{El-Haj:2025zbe,Philcox:2024jpd}. 

From Sec.~VIII of~\cite{El-Haj:2025zbe}, the contribution of a single hotspot at position $\mathbf{x}_{\rm HS}$ to the harmonic-space CMB fluctuation field ($X \in [T,E]$) is given by
\begin{align}\label{eq: cmb-effect}
    \delta X_{\ell m}(\mathbf{x}_{\rm HS}) &= \frac{2}{\pi}\int\frac{\dd k}{k}\mathcal{T}^X_\ell(k)\frac{gH_I^2}{\dot{\phi}_0}\left(\text{Si}(k\eta_*)-\sin(k\eta_*)\right)\nonumber\\
    &\quad\,\times\,i^\ell  \int\frac{\dd\hat{\mathbf{k}}}{4\pi}e^{-i\mathbf{k}\cdot\mathbf{x}_{\rm HS}}Y^*_{\ell m}(\hat{\mathbf{k}}),
\end{align}
where the second line evaluates to $Y_{\ell m}^*(\hat{\mathbf{n}}_{\rm HS})j_\ell(k\chi_{\rm HS})$ if we write $\mathbf{x}_{\rm HS}=(\eta_0-\eta_{\rm HS})\hat{\mathbf{n}}_{\rm HS}$.
To compute the one-hotspot term in our model, we simply square this and average over $\mathbf{x}_{\rm HS}$, finding~\cite{El-Haj:2025zbe}:
\begin{widetext}
    \begin{align}
    \label{eq.Cl1h}
C_\ell^{XY,1\text{h}} &\equiv n_{\rm co}\int \frac{\dd^3\mathbf{x}_{\rm HS}}{\mathcal{V}_{\rm HS}}\,\langle \delta X_{\ell m}(\mathbf{x}_{\rm HS})\delta Y^*_{\ell m}(\mathbf{x}_{\rm HS})\rangle\nonumber \\\nonumber
    &=\frac{2 n_{\text{co}}}{\pi}\int \frac{\dd k}{k^4} \mathcal{T}^X_\ell(k)\mathcal{T}^Y_\ell(k)\left(\frac{gH_I^2}{\dot{\phi}_0}\right)^2\left(\text{Si}(k\eta_*)-\sin(k\eta_*)\right)^2\nonumber \\
    &\equiv\frac{2}{\pi}\int \dd k k^2\mathcal{T}_\ell^X(k)\mathcal{T}^Y_\ell(k) \left(2g^2P_*P_{\rm{std}}(k)\times \frac{\left(\text{Si}(k\eta_*)-\sin(k\eta_*)\right)^2}{(k\eta_*)^3}\right) \,. 
\end{align}
\end{widetext}
Here, we have used the orthogonality of the spherical Bessel functions and assumed that the particles are uniformly distributed throughout the Hubble volume. Notably, this matches the one-hotspot term of the primordial derivation.

To build the two-hotspot term, we correlate a pair of CMB fluctuations sourced by particles at $\mathbf{x}_{\rm HS}$ and $\mathbf{x}_{\rm HS}+\mathbf{d}$. Averaging over $\mathbf{x}_{\rm HS}$ and $\mathbf{d}$ (with volumes $\mathcal{V}_{\rm HS}$ and $\mathcal{V}_d$), we find
\begin{widetext}
    \begin{align}
    \label{eq.Cl2h}
C_\ell^{XY,2\text{h}} &\equiv n_{\rm co}\int \frac{\dd^3\mathbf{x}_{\rm HS}}{\mathcal{V}_{\rm HS}}\frac{\dd^3\mathbf{d}}{\mathcal{V}_d}\,\langle \delta Y_{\ell m}(\mathbf{x}_{\rm HS}+\mathbf{d})\delta X^*_{\ell m}(\mathbf{x}_{\rm HS})\rangle\\\nonumber
    &=  \frac{2n_{\rm co}}{\pi}\int\frac{\dd k}{k^4}\mathcal{T}^X_\ell(k)\mathcal{T}^Y_\ell(k)\left(\frac{gH_I^2}{\dot{\phi}_0}\right)^2\left(\text{Si}(k\eta_*)-\sin{k\eta_*}\right)^2\int \frac{\mathrm{d}\mathbf{d}}{\mathcal{V}_d}\mathrm{d}\hat{\mathbf{k}}\,e^{-i\mathbf{k}\cdot\mathbf{d}}Y^*_{\ell m}(\hat{\mathbf{k}})Y_{\ell m}(\hat{\mathbf{k}})\\\nonumber
    &=  \frac{2n_{\rm co}}{\pi}\int\frac{\dd k}{k^4}\mathcal{T}^X_\ell(k)\mathcal{T}^Y_\ell(k)\left(\frac{gH_I^2}{\dot{\phi}_0}\right)^2\left(\text{Si}(k\eta_*)-\sin{k\eta_*}\right)^2\left(3\frac{\sin{k\eta_*}-k\eta_*\cos{k\eta_*}}{(k\eta_*)^3}\right)\\
    &\equiv\frac{2}{\pi}\int \dd k k^2\mathcal{T}_\ell^X(k)\mathcal{T}^Y_\ell(k) \left(6g^2P_*P_{\rm{std}}(k)\times \frac{\left(\text{Si}(k\eta_*)-\sin{k\eta_*}\right)^2}{(k\eta_*)^3}\times\frac{\sin{k\eta_*-k\eta_*\cos{k\eta_*}}}{(k\eta_*)^3}\right) \,.
\end{align}
\end{widetext}
where the final line follows from assuming a uniform distribution for $|\mathbf{d}|\in[0,\eta_\star]$.
As expected, this ``late time'' computation consistently reproduces the result from the primordial calculation. 

\subsection{Analogy with the Halo Model}

\noindent Here we briefly note that there is an exact equivalence between the formalism underlying the hotspot contributions to the CMB and the halo model frequently used to describe large-scale structure statistics and CMB secondary anisotropies~\cite[e.g.,][]{Seljak:2000gq,Cooray:2002dia}.  A brief dictionary relating the two pictures is as follows.  Each hotspot is equivalent to a ``halo'' with a given profile --- in the hotspot case, the profile is most naturally expressed in the primordial curvature perturbation field, $f(k\eta_*) \equiv \frac{gH_I^2}{\dot{\phi}_0}(\text{Si}(k\eta_*)-\sin{k\eta_*})$.  One can then construct the associated $T$ or $E$ profile, cf.~\autoref{eq: cmb-effect}.  The analog of the halo mass function is the hotspot number density, cf.~\autoref{eqn:number}.  One key difference is that in our model, all of the hotspots are produced at a single epoch and hence have a single characteristic size $\eta_*$, whereas in the halo model there are halos of a wide variety of masses and redshifts.\footnote{This property of our inflationary model can easily be generalized to encompass models with repeated bursts of particle production, as in~\cite{Flauger_2017,M_nchmeyer_2019}.}

As in the halo model, the one-hotspot (one-halo) term is given by integrating over the hotspot number density weighted by the squared profile (\autoref{eq.Cl1h}).  There is also a ``two-halo'' clustering contribution, which arises because the hotspots are produced in pairs (two hotspots per particle-production event).  However, the events are assumed to be Poisson-distributed (randomly within the Hubble volume at the production epoch) and thus there are no inter-event correlations, only correlations between the two hotspots in a given pair.  There is hence no non-trivial power spectrum associated with the underlying particle-production field; the hotspots are not ``biased tracers'' of an underlying distribution.  The two-hotspot (two-halo) term is therefore simpler in some sense in our model than in the halo model, with the clustering contribution determined by placing the particle-pairs uniformly within the Hubble volume at production, leading to \autoref{eq.Cl2h}.  Other results in the halo model, e.g., higher-point correlation functions or the one-point PDF, also have direct analogs in the context of our hotspot model, which we leave to future work.


\section{\label{sec:data} Data}
\noindent 

\noindent We use the latest CMB measurements from ACT DR6, both alone and in combination with the large-scale \emph{Planck} measurements as prescribed in the \texttt{P-ACT} data combination~\cite{ACT:2025fju,ACT:2025tim}. This combination uses the full ACT DR6 power spectra in temperature and polarization from $\ell = 600$ to $6500$, combined with the \emph{Planck} PR3 spectra restricted to $\ell < 1000/600/600$ in TT/TE/EE~\cite{Planck2018Likelihood}. We also use the large-scale temperature power spectrum from the \texttt{Commander} pipeline, and the large-scale E-mode measurements from the \texttt{Sroll2} likelihood~\cite{Sroll2}. We use the publicly available \texttt{DR6-ACT-lite} pre-marginalized likelihoods and data, which includes the likelihood for the pre-marginalized \emph{Planck} data. To get a measure on the constraints coming from the individual datasets and their union, we compare the constraints from \emph{Planck}-only data, ACT-only data, and the \texttt{P-ACT} combination. We finally investigate the constraints coming from a joint analysis with the lensing of the CMB as measured by the combination of ACT DR6 and \emph{Planck} PR4~\cite{Qu_2024,Madhavacheril_2024,Carron_2022}, and measurements of baryon acoustic oscillations from DESI DR2~\cite{DESI:2025zgx,DESI:2025zpo}, in a combination known as \texttt{P-ACT-LB}\footnote{It should be noted that \texttt{P-ACT-LB} in \cite{ACT:2025fju} refers to using DESI DR1, while here we use the newer DESI DR2 dataset.}.

\section{Methodology}
\label{sec:sims}
\noindent 

\noindent We model the modified primordial power spectrum using the closed-form parametric expression in~\autoref{eq:pps-full-expression}, implementing it as a modified primordial power spectrum in the Einstein-Boltzmann code \texttt{camb}~\cite{Lewis:1999bs},\footnote{\url{https://camb.info/}} using the standard power-law expression $P_{\rm std}(k) \propto k^{n_s-1}$. We use the \texttt{camb} accuracy settings defined in Appendix A of Ref.~\cite{ACT:2025tim}.

We show the CMB power spectra from $\Lambda$CDM and modified PPS models in~\autoref{fig:pps-fullcamb}, and perform a comparison between the linear and non-linear matter power spectra in~\autoref{fig:matter-power}. We note that we use the non-linear model from \texttt{mead2020} in \texttt{camb}, which predicts the effects of nonlinear structure formation and baryonic feedback based around a fiducial $\Lambda$CDM model, and might not be accurate for the deviations caused by our primoridial hotspot model. These effects could couple to the small-scale CMB anisotropies through the lensing-induced imprint of the matter power spectrum. However, we expect that the deviations from the nonlinear model will be small corrections on top of those from the modified PPS, and thus we leave detailed consideration of this effect to future work.

\subsection{Bayesian Inference}

\noindent We perform Bayesian inference using the Markov Chain Monte-Carlo (MCMC) package \texttt{Cobaya}. The aforementioned datasets each have their own interfaces with \texttt{Cobaya}, facilitating straightforward analyses of multiple dataset combinations over the same parameter ranges. We impose uninformative flat priors on the six $\Lambda$CDM parameters $\Omega_b h^2$, $\Omega_c h^2$, $\theta_{\rm MC}$, $n_s$, $\log A_s$, $\tau_{\rm reio}$. Since the modified power spectrum is parameterized by the combination $2 g^2 P_*$, we choose to impose upon this combined amplitude parameter a flat, uninformative positivity prior $2 g^2 P_* \ge 0$. This option gives us information about the amplitudes allowed by our data. We finally keep the parameter $\log_{10} \eta_*$ fixed, and infer the constraints on the amplitude as a function of the scale by running separate MCMC analyses for different values of $\eta_*$.  This mitigates any ``prior volume effects'' in the Bayesian posteriors.


\subsection{Fisher Forecast}

\noindent To supplement our observational results, we perform a Fisher estimate to forecast the constraining power of a cosmic variance-limited (CVL) experiment, and a Simons Observatory-like (SO-like) experiment. We build a covariance matrix of a CMB experiment that measures the TT, TE, and EE power spectrum between $2 \le \ell \le 8500$, assuming the covariance between the XY and ZW spectra is
\begin{widetext}
\begin{equation}
    \Sigma_{\ell, \ell'}^{XY,ZW} = \frac{\delta_{\ell,\ell'}}{f_{\rm sky}(2 \ell + 1)} \left[ \left(C_\ell^{XZ} + N_\ell^{XZ}\right) \left(C_\ell^{YW} + N_\ell^{YW}\right) + \left(C_\ell^{XW} + N_\ell^{XW}\right) \left(C_\ell^{YZ} + N_\ell^{YZ}\right) \right]
\end{equation}
\end{widetext}
where $f_{\rm sky}$ is the sky fraction of the experiment, which is $100\%$ ($40\%$) for CVL (SO-like), and $N_\ell^{XY}$ is the $XY$ noise spectrum, which for CVL we set to zero, while for SO-like we use the public SO noise spectra\footnote{Available at \url{https://github.com/simonsobs/so_noise_models}.}.

For each fixed value of $\eta_*$, we obtain the power spectrum around $2 g^2 P_* = 0$, using the best-fit \texttt{P-ACT} values for $\Lambda$CDM parameters. We compute the Fisher information matrix as:
\begin{equation}
    F_{ij} = \sum\limits_{XY,ZW,\ell_1,\ell_2} \frac{\partial C^{XY}_{\ell_1}}{\partial \theta_i} \left( \Sigma_{\ell_1, \ell_2}^{XY,ZW} \right)^{-1} \frac{\partial C^{ZW}_{\ell_2}}{\partial \theta_j} ,
\end{equation}
where $F_{ij}$ contains the Fisher information about the $\theta_i,\theta_j$ parameter covariance. In order to improve numerical stability of the matrix inversion and computational speed, we employ a flat $\Delta \ell = 30$ binning across the power spectra. We estimate the derivatives by taking finite differences over the individual parameters, varying them by $0.1\%$.

We compute the seven-dimensional Fisher information matrix for the six $\Lambda$CDM parameters plus the amplitude $2 g^2 P_*$. We then compute the Fisher forecasted error on parameter $i$ as:
\begin{equation}
    \sigma_i = \sqrt{ (F^{-1})_{ii} } .
\end{equation}
It should be noted that the likelihoods for the angular power spectra for the $2 \le \ell \lesssim 30$ modes are non-Gaussian, and our flat binning and Gaussian treatment of them might not be accurate. In principle however, these modes only help constrain the optical depth at reionization $\tau_{\rm reio}$ (at least for EE), and we find that this parameter is minimally correlated with our parameter of interest $2 g^2 P_*$. We verify that starting at $\ell > 30$, or not binning the $\ell < 30$ modes, provides similar results to those presented here.

\section{Results}
\label{sec:Results}
\noindent 

\subsection{Primordial Power Spectrum}

\begin{figure}
    \includegraphics[width=\linewidth]{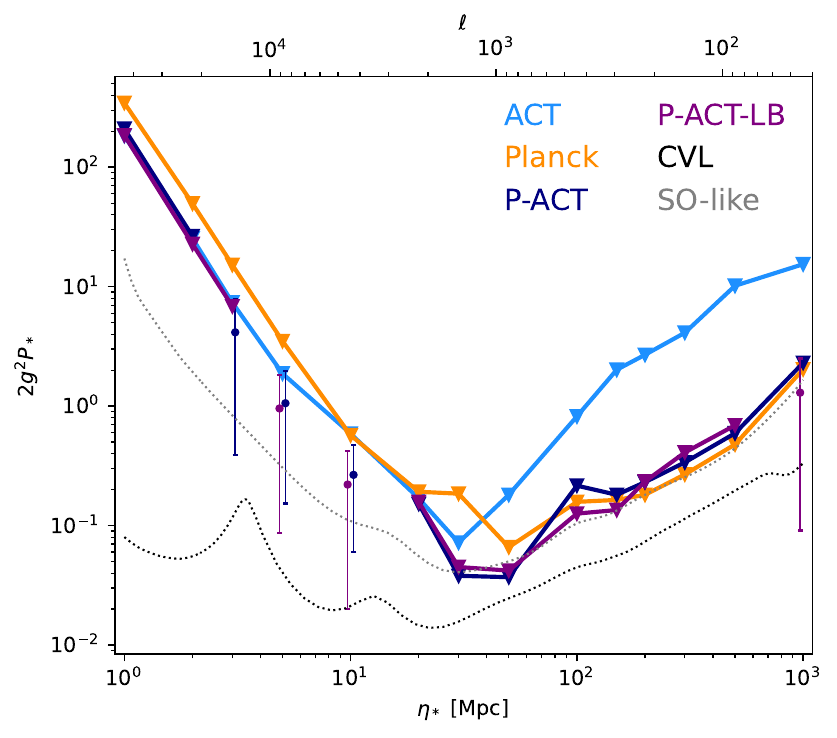}
    \caption{$95\%$ upper limits on the massive-particle power spectrum contribution for fixed size parameter $\eta_*$. The top axis refers to the approximate $\ell$ mode where the hotspot power spectrum contribution is maximized. \emph{Planck} (\textbf{orange}) dominates the constraints at $\eta_* > 30 \, \rm{Mpc}$, while ACT (\textbf{light blue}) leads for $\eta_* \le 30 \, \rm{Mpc}$. The combination of the two, in \texttt{P-ACT} (\textbf{navy}) and in combination with lensing and DESI (\textbf{purple}), provides the tightest constraints in the range $10 \, \rm{Mpc} \le \eta_* \le 100 \, \rm{Mpc}$. The circular data points refer to the $2 \sigma$ intervals of the constraints where the lower limit is nonzero (cf.~\autoref{tab:pps-upper-limits-values}, the points are offset horizontally for readability). The dotted \textbf{black} and \textbf{grey} lines show the Fisher forecast for the constraining power of a cosmic variance-limited and SO-like experiment, respectively.}
    \label{fig:pps-upper-limits-grid}
\end{figure}

\noindent We obtain the posterior density distribution in the hotspot amplitude-size parameter plane, $(2 g^2 P_*, \eta_*)$, taking 14 logarithmically-spaced steps between $1 \, \rm{Mpc} \le \eta_* \le 1000 \, \rm{Mpc}$, and imposing a flat positivity prior on the amplitude $2 g^2 P_*$. We find no significant preference ($\geq 3\sigma$) for a hotspot contribution to the primordial power spectrum at any length scale. We show the upper limits on the primordial power spectrum contribution in~\autoref{fig:pps-upper-limits-grid}. In addition, we provide the numerical values of our constraints in~\autoref{tab:pps-upper-limits-values}.

We find that the \emph{Planck} data dominate the constraints for hotspots with characteristic scales $\eta_* > 30 \, \rm{Mpc}$, while the ACT data improve the constraints at $\eta_* \le 30 \, \rm{Mpc}$. The inclusion of the ACT data tightens the bounds on these scales by approximately a factor of two.
Even though ACT does not contain data below $\ell < 600$, and \textit{Planck} does not contain data above $\ell > 2500$, the two datasets can still place loose constraints on the large-$\eta$ and small-$\eta$ contributions respectively, due to the $k^{-3}$ tails of the primordial template --- see~\autoref{fig:pps-modified}.

\begin{table}
    \centering
    \begin{tabular}{c|c c c c}
        \toprule
        $\eta_*$ & \emph{Planck} & ACT & \texttt{P-ACT} & \texttt{P-ACT-LB} \\
        \midrule
        $1 \, \mathrm{Mpc}$ & $< 345$ & $< 182$ & $< 209$ & $< 184$ \\
        $2 \, \mathrm{Mpc}$ & $< 49.7$ & $< 25.4$ & $< 26.4$ & $< 22.7$ \\
        $3 \, \mathrm{Mpc}$ & $< 15.2$ & $< 7.37$ & $4.3 \pm 2.0$ & $< 6.9$ \\
        $5 \, \mathrm{Mpc}$ & $< 3.50$ & $< 1.87$ & $1.08 \pm 0.47$ & $0.99^{+0.44}_{-0.51}$ \\
        $10 \, \mathrm{Mpc}$ & $< 0.57$ & $< 0.59$ & $0.26 \pm 0.11$ & $0.23^{+0.10}_{-0.12}$ \\
        $20 \, \mathrm{Mpc}$ & $< 0.19$ & $< 0.17$ & $< 0.15$ & $< 0.16$ \\
        $30 \, \mathrm{Mpc}$ & $< 0.19$ & $< 0.072$ & $< 0.038$ & $< 0.045$ \\
        $50 \, \mathrm{Mpc}$ & $< 0.66$ & $< 0.181$ & $< 0.037$ & $< 0.042$ \\
        $100 \, \mathrm{Mpc}$ & $< 0.16$ & $< 0.820$ & $< 0.22$ & $< 0.13$ \\
        $150 \, \mathrm{Mpc}$ & $< 0.16$ & $< 2.02$ & $< 0.18$ & $< 0.14$ \\
        $200 \, \mathrm{Mpc}$ & $< 0.18$ & $< 2.69$ & $< 0.23$ & $< 0.23$ \\
        $300 \, \mathrm{Mpc}$ & $< 0.27$ & $< 4.11$ & $< 0.34$ & $< 0.41$ \\
        $500 \, \mathrm{Mpc}$ & $< 0.48$ & $< 10.2$ & $< 0.59$ & $< 0.69$ \\
        $1000 \, \mathrm{Mpc}$ & $< 2.03$ & $< 15.4$ & $< 2.30$ & $1.35^{+0.60}_{-0.69}$ \\
        \bottomrule
    \end{tabular}
    \caption{Constraints on $2 g^2 P_*$ from various data set combinations. The bounds given are the $95\%$ C.L. upper limit for values consistent with zero, or the $1 \sigma$ bounds for nonzero values. We find only a mild preference for a nonzero amplitude in the range $3 -10\,\mathrm{Mpc}$ when combining ACT and \emph{Planck} data, although this preference is slightly reduced through the inclusion of lensing and BAO data.}
    \label{tab:pps-upper-limits-values}
\end{table}

We do not find a preference for a non-zero hotspot amplitude in either ACT or \emph{Planck} data. The combination of the two tightens the upper limit across the entire range, and additionally yields a mild $2.2-2.6 \sigma$ hint for $2g^2P_*>0$ in the range $3-10 \, \rm{Mpc}$. This hint is slightly reduced with the inclusion of lensing and BAO data, although it remains just above the $2 \sigma$ level. At this scale size, the feature should peak around $\ell \sim 4500-15000$, which is on the edge of the ACT data.  We also emphasize that the significance values quoted here are local assessments only; no global ``look-elsewhere'' penalty is included.  Thus, even if real, the true significance of the feature would be smaller than the values quoted here.


There has been interest recently in the value of $n_s$, as the inclusion of small-scale data from ACT slightly increases the value of $n_s$ by $\sim 1 \sigma$, from $\approx 0.965$ to $\approx 0.970$ when combined with \emph{Planck}\footnote{Quantitatively, in $\Lambda$CDM, $n_s = 0.9651 \pm 0.0044$ for \emph{Planck} alone and $n_s = 0.9709 \pm 0.0038$ for \texttt{P-ACT}~\cite{ACT:2025fju}.} \cite[see, e.g.,][]{choudhury2025quintessentialinflationlightact,mcdonough2026spectrumnsconstraintsdesi,bezerrasobrinho2025starobinskyinflationlatestcmb,maity2026actinginflationimplicationsnon,haque2025minimalplateauinflationlight,Okada_2025,lynker2025actimplicationshilltopinflation,Zharov_2025}.  Inclusion of DESI BAO data leads to a further $\sim 1 \sigma$ increase, to $\approx 0.975$~\cite{ACT:2025fju}.  Hotspots around the size $\eta_* \lesssim 10 \, \mathrm{Mpc}$ would induce features in the primordial power spectrum above the pivot scale $k_0 = 0.05 \, \mathrm{Mpc}^{-1}$, and could thus account for excess power in the primordial power spectrum on smaller scales. It is possible that our slight preference for a primordial feature around this scale relates to the mild increase in the value of $n_s$.

\begin{figure*}
    \centering
    \includegraphics[width=\linewidth]{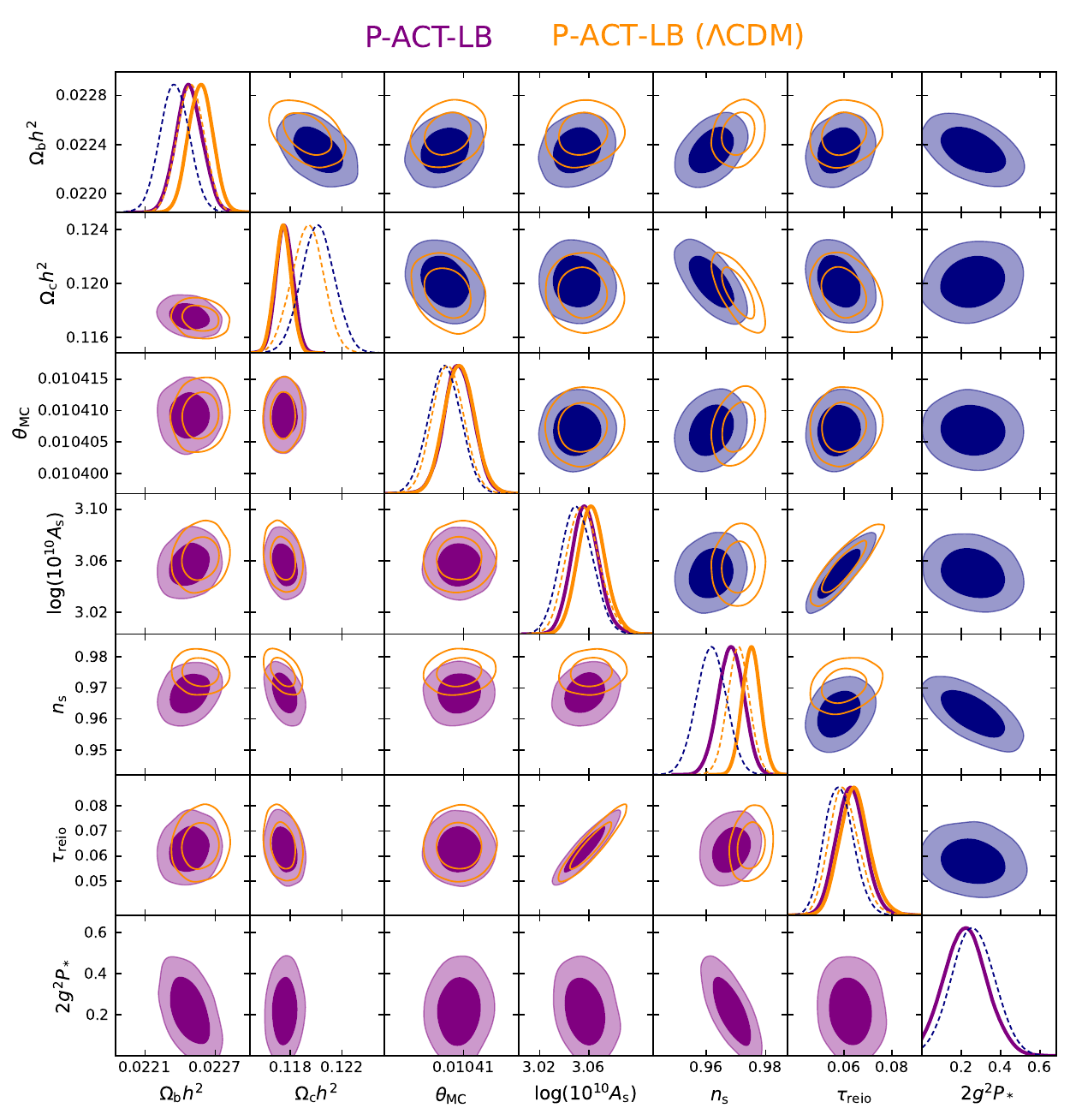}
    \caption{Constraints on the six $\Lambda$CDM cosmological parameters from \texttt{P-ACT} and \texttt{P-ACT-LB} for both $\Lambda$CDM and the modified primordial-power-spectrum model, for a massive particle feature with size $\eta_* = 10 \, \mathrm{Mpc}$. The lower triangle shows the constraints from \texttt{P-ACT-LB}, with the \textbf{purple} contours including the hotspot contribution and the \textbf{orange} contours showing the $\Lambda$CDM constraints. The upper triangle shows the same for \texttt{P-ACT} in \textbf{navy} and \textbf{orange} respectively, with these \texttt{P-ACT} constraints being shown as \textbf{dashed} lines along the diagonal. 
    }
    \label{fig:double-triangle-10}
\end{figure*}

We show the constraints on the six $\Lambda$CDM cosmological parameters in~\autoref{fig:double-triangle-10}, comparing both the $\Lambda$CDM bounds and the extension with a hotspot contribution at $\eta_* = 10 \, \mathrm{Mpc}$, using both \texttt{P-ACT} and \texttt{P-ACT-LB}. For the most part, the background cosmology is stable when opening up this extra parameter, with only a minor shift in the $\Omega_b h^2$ -- $n_s$ plane. This can be explained simply by the correlation between the two parameters within the CMB power spectrum, with the spectral index $n_s$ being given more freedom when the hotspot feature is included. The remaining four parameters are stable against the inclusion of our feature. Note that the inclusion of this feature correlates with lower values of $n_s$, as the excess power created would accommodate a steeper primordial power spectrum, hence the value for $n_s$ is allowed to drop back to $n_s \approx 0.965$. We show the effect of including hotspots on the inferred cosmological parameters for different $\eta_*$ values in Appendix~\ref{app:cosmo-figures}.

\subsection{Interaction Parameter Space}

\begin{figure*}
    \includegraphics[width=0.45\linewidth]{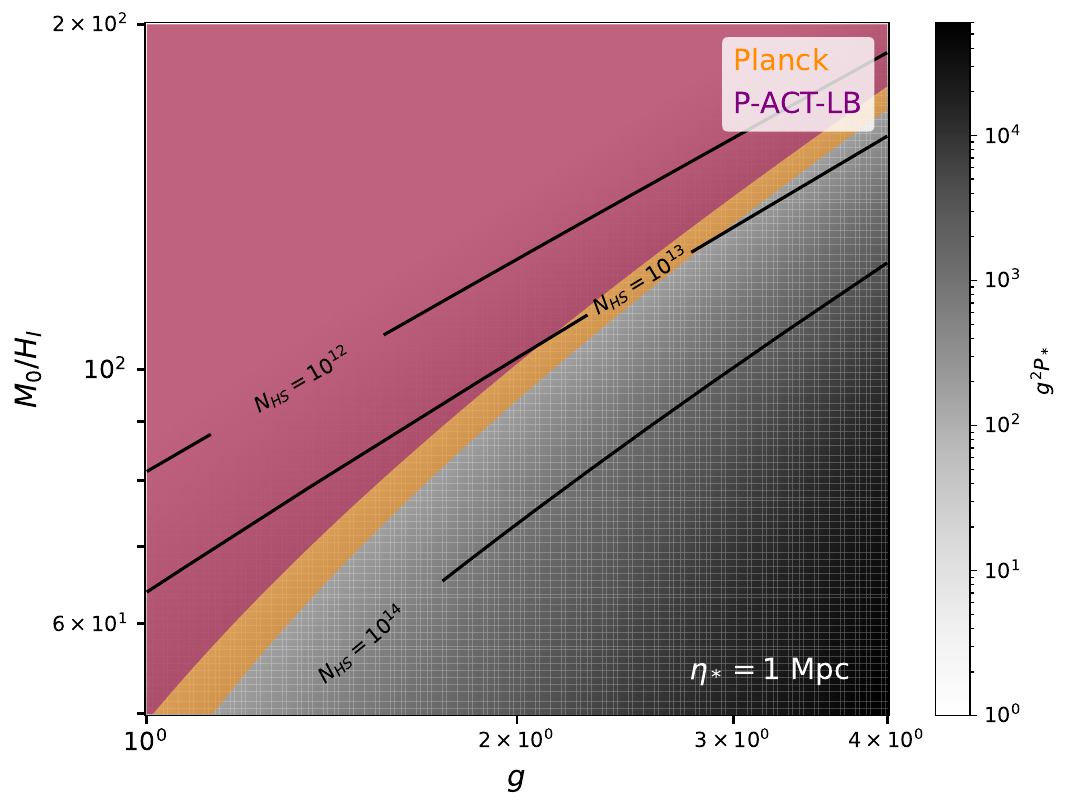} \includegraphics[width=0.45\linewidth]{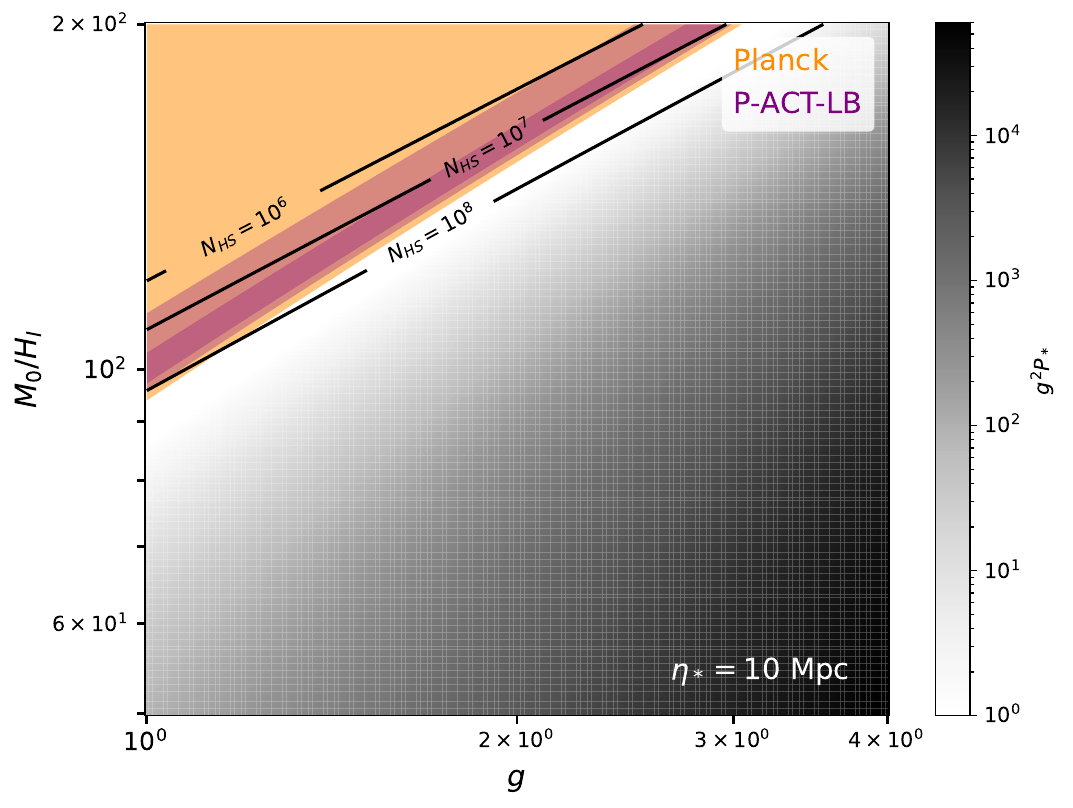}
    \includegraphics[width=0.45\linewidth]{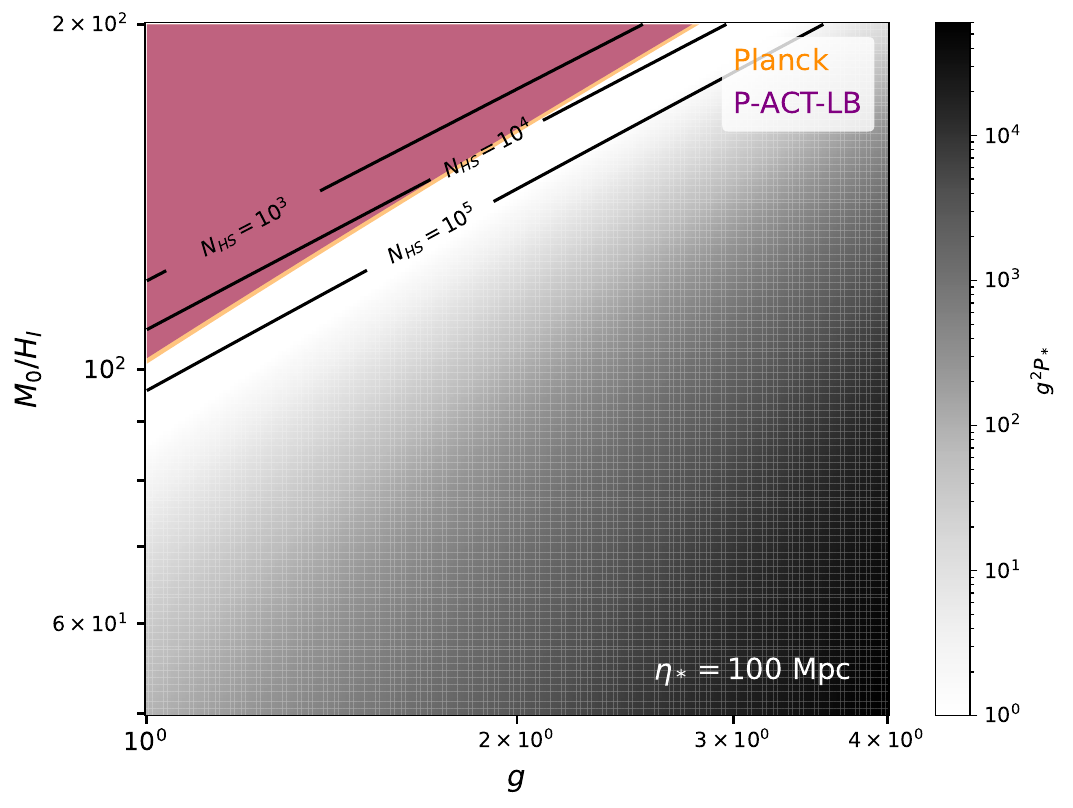} \includegraphics[width=0.45\linewidth]{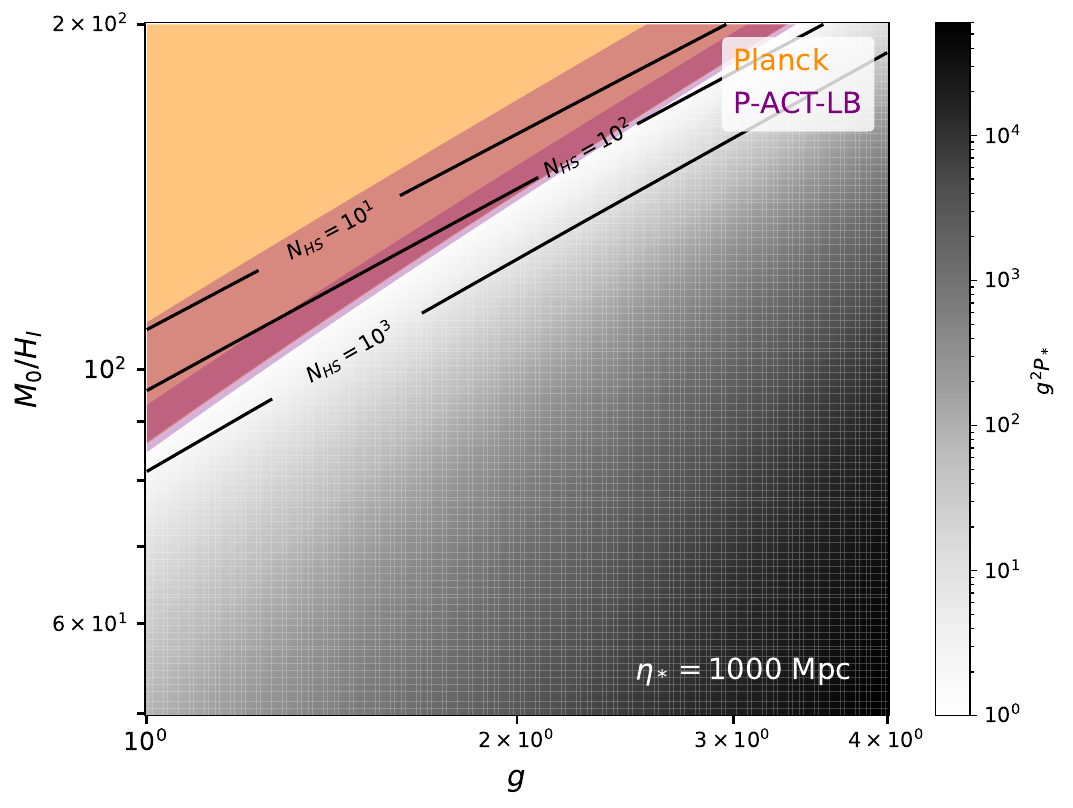}
    \caption{Constraints on the coupling constant $g$ and bare mass $m_0$ from \emph{Planck} (\textbf{orange}) and \texttt{P-ACT-LB} (\textbf{purple}). The colored regions indicate the parameter space allowed/preferred by the $g,m_0$ combination (i.e., regions to the right/bottom are ruled out). The grayscale background shows the amplitude $g^2 P_*$ as a function of $g, m_0$. The black isocontours show lines of constant numbers of hotspots generated ($N_{\rm HS}$) for fixed $\Delta \eta = 100 \, \mathrm{Mpc}$. The different panels refer to different size scales, ranging from the smallest features in the top-left, to the largest features in the lower right. We see that \texttt{P-ACT-LB} considerably improves on the bounds from \emph{Planck}, especially on smaller scales.}   
    \label{fig:int-params-constraints}
\end{figure*}

\noindent By inverting~\autoref{eq:pps-amplitude}, we can translate the upper limit on $2 g^2 P_*$ into a constraint on the massive-particle interaction parameter space. Because of the degeneracy between the coupling $g$ and the mass parameter $m_0$, the constraint on $2 g^2 P_*$ excludes a region in the $(g,m_0)$ plane for a fixed $\eta_*$. As an example, we show the constraints for $\eta_* = 10\, \rm{Mpc}$ in~\autoref{fig:int-params-constraints} for both \emph{Planck} and \texttt{P-ACT-LB}.

Comparing with the values from~\autoref{tab:pps-upper-limits-values}, we see that the slight preference for the non-zero amplitude at $\eta_* = 10 \, \mathrm{Mpc}$ and $\eta_* = 1000 \, \mathrm{Mpc}$ in \texttt{P-ACT-LB} translates into a band of values preferred by this dataset combination  in~\autoref{fig:int-params-constraints}. Despite this, the $95 \%$ upper bound does move towards the top-left corner of this parameter space, as the total upper bound on $2 g^2 P_*$ does lower with the inclusion of small-scale data. We should note that the number of expected produced particles $\sim\frac{1}{\eta_*^3}$\cite{El-Haj:2025zbe,Philcox:2024jpd,M_nchmeyer_2019}, and as such the strongest preferences (and the regimes in which ACT greatly tightens the constraints) are in a regime where particle production is most numerous, and the power spectrum is expected to be the most constraining statistic. Note that once again, we see from \autoref{fig:int-params-constraints} that we are able to constrain perturbative values for the coupling (naively $g\lesssim \sqrt{4\pi}$), as well as masses $m_0=\mathcal{O}(100H_I)$, for which the theory detailed in \autoref{sec:theory}
is self-consistent. We should also note that these are energy scales vastly higher than those achievable at a terrestrial collider.  Coupled with the results of \cite{Philcox:2024jpd,El-Haj:2025zbe}, these constraints represent some of the highest energies observationally probed.

\subsection{Comparison With Matched-Filter Approach}

\noindent It is instructive to compare our constraints to those obtained using the matched-filter search described in \cite{Philcox:2024jpd,El-Haj:2025zbe}. Here, we briefly summarize the matched-filter formalism of \citep{El-Haj:2025zbe}, and forecast results from both approaches\footnote{We ignore the effects of `confusion noise', arising from the regime in which the hotspots cover a non-negligible fraction of the sky and thus induce intrinsic Poissonian noise and a non-Gaussian covariance contribution (as in halo model calculations). We do not expect this to play a large role in our constraints given the tight upper limits already found on the hotspot contribution, which imply that noise from the $\Lambda$CDM fluctuations and instrument noise are dominant.} 
for a cosmic-variance-limited experiment.

Models of the form studied in this work induce localized position-space features in the CMB, which in our case take the form (following \autoref{eq: cmb-effect})~\cite{Kim:2021ida},
\begin{align}
    \label{eq:positionspace}
    \delta X(\hat{\bf{n}},\hat{\bf{n}}_{{\mathrm{HS}}},\eta_*,\eta_{{\mathrm{HS}}})&=\frac{1}{2\pi^2}\sum_{\ell=2}^\infty (2\ell+1)\mathcal{P}_\ell(\vb{\hat{\bf{n}}}\vdot\vb{\hat{\bf{n}}_{{\mathrm{HS}}}})\nonumber\\
    &\times\int\frac{\dd k}{k}\frac{gH_I^2}{\dot\phi_0}\left(\mathrm{Si}(k\eta_*)-\sin(k\eta_*)\right) \nonumber\\
    &\,\times\,j_\ell(k\chi_{{\mathrm{HS}}})\mathcal{T}^X_\ell(k)\,,
\end{align}
where $\mathcal{P}_\ell$ are Legendre polynomials. Noticing that the profiles are linear in $g$, one can construct an inverse-variance-weighted estimator for $g$:
\begin{align}
       \widehat{g}(\boldsymbol{\theta})&=\sigma_g^2\int \frac{\dd ^2\vb{l}}{(2\pi)^2}\frac{(t^X)^*(\vb{l},\boldsymbol{\theta})d^X(\vb{l})}{C^{XX}(\vb{l})} \,,\\
        \sigma_g^{X}&=\left( \int \frac{\dd^2 \vb{l}}{(2\pi)^2}\frac{\abs{t^X(\vb{l,\vb{0}})}^2}{C^{XX}(\vb{l})} \right)^{-\frac{1}{2}} \,,
        \label{eq:MF}
\end{align}
where, $t^X(\vb{l},{\boldsymbol \theta})$ is the Fourier-space $X$-mode hotspot profile centered at ${\boldsymbol \theta}$, and $d^X({\boldsymbol\theta})$ is a component-separated CMB map for $X\in\{T,E\}$ and we have adopted the flat-sky limit for simplicity. The power spectrum $C^{XX}(\vb{l})$ includes both signal and experimental noise, and is directly computed from the data. For actual searches, the sky is tiled, and the estimator is computed on a tile-by-tile basis.
The matched filter search identifies hotspot candidates as regions with $\text{SNR}=\widehat{g}/\sigma_g\geq 5$. Critically, no strong candidates were found in the searches conducted in Refs.~\cite{El-Haj:2025zbe,Philcox:2024jpd}. 

To place bounds on the underlying physics, we build a Poissonian likelihood \cite{2001ApJ...560L.111H}, which in the limit where no hotspots are observed reduces to
\begin{align}\label{eq:likelihood_simple}
    \ln \mathcal{L}(g,\eta_*,m_0) = -N_{\text{pred}}(g,\eta_*,m_0) \,.
\end{align}
The predicted number of hotspots is given by
\begin{align}\label{eq:Npred}
    N_{\text{pred}}(g,m_0,\eta_*) = \left(\frac{1}{H_I\eta_*}\right)^3 \int \dd^3 x \, n(g,m_0) \,\mathcal{S}(g,\eta_*,\eta_{\text{HS}}) \,,
\end{align}
where the number density is given by \autoref{eqn:number}, and $\mathcal{S}(g,\eta_*,\eta_{\text{HS}})$ is the selection function, \textit{i.e.}, the probability of detecting an individual hotspot with parameters $(g,\eta_*,\eta_{\text{HS}})$, which is derived in \cite{El-Haj:2025zbe}. We compute $\sigma_g$ and the selection function analytically for computational ease (rather than on a tile-by-tile basis in the \emph{Planck} data). Fixing to characteristic values of $m_0$, one can perform one-parameter fits for a variety of $\eta_*$ values (here we present 10 logarithmically spaced $\eta_*$ values). 

Note that the matched filter and power spectrum naturally constrain different  parameters in this search. The matched filter technique constrains the amplitude of the position-space profile ($g$), which is independent of the number of produced hotspots; the predicted number is included only later when we derive bounds via the Poissonian likelihood. This hints at the fact that this single-matched-filter is only the optimal statistic in the theory space where a small number of particles are produced. In fact, though currently computationally intractable, the optimal statistic would be an $N_{\rm{HS}}-$point matched filter. By contrast, the power spectrum search probes directly the amplitude of the feature, which is a degenerate combination of $g,m_0$ that determines the number of produced particles. As such, to compare the two approaches, we must fix the values of $m_0$ considered for the matched filter.  

\begin{figure*}[t]
\centering
    \includegraphics[width=0.48\textwidth]{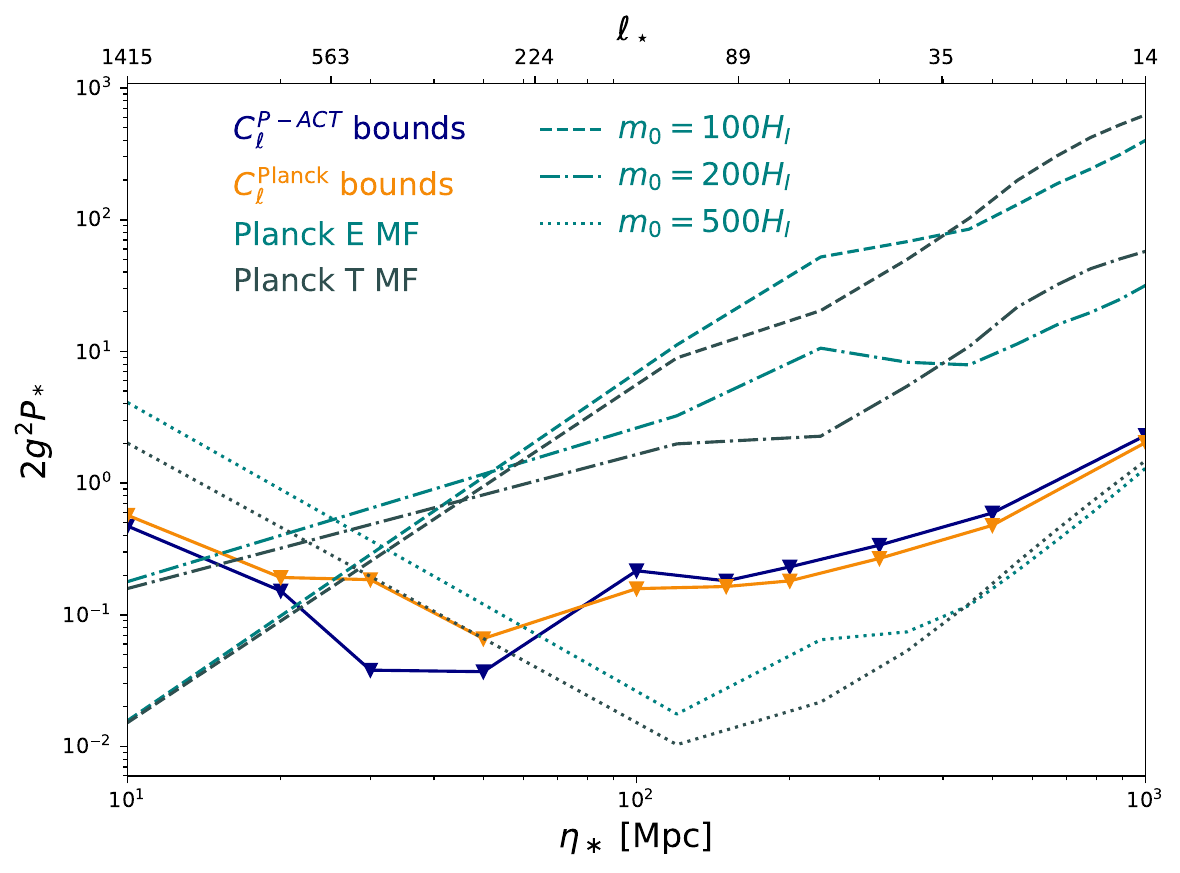}
  \hfill
  \includegraphics[width=0.48\textwidth]{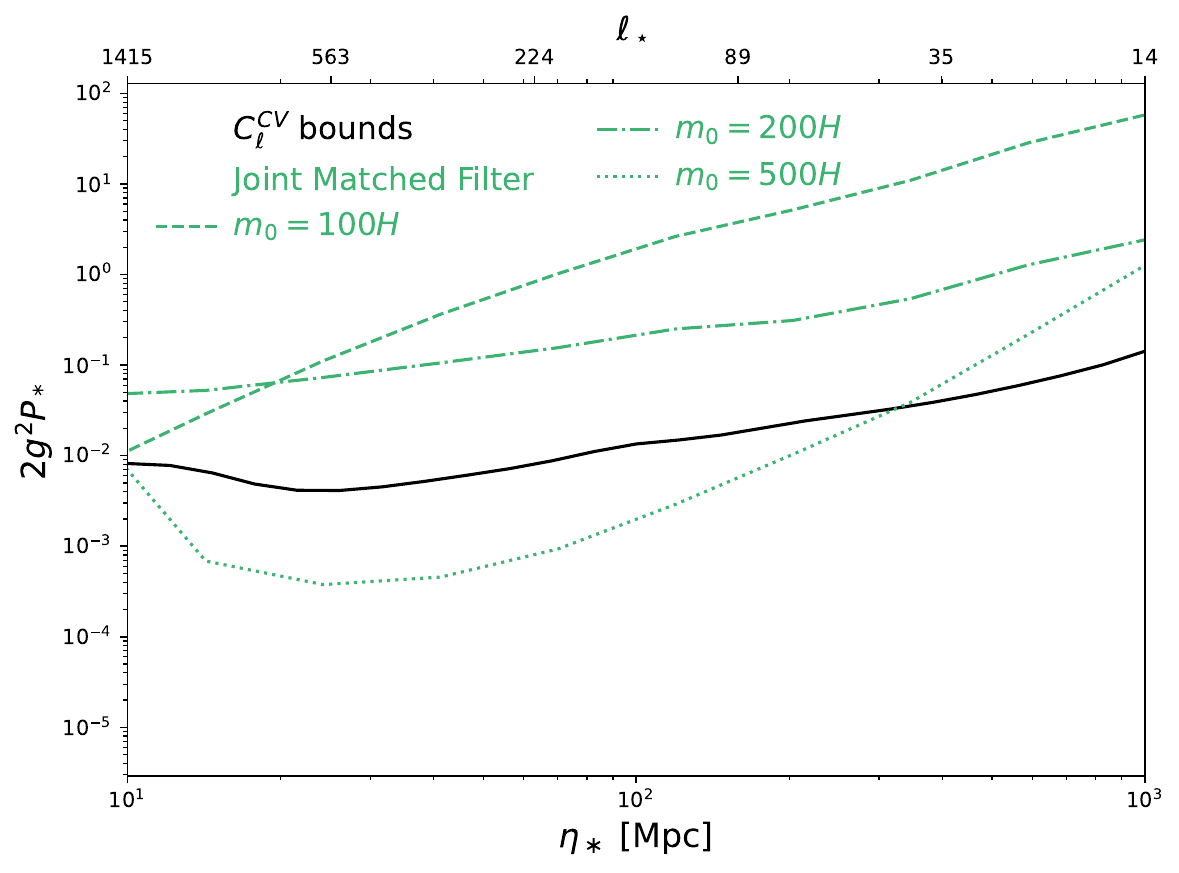}
  \caption{A comparison of the hotspot constraints derived from the power spectrum analysis in this work to those from the localized matched-filter analyses of \cite{El-Haj:2025zbe,Philcox:2024jpd}. 
  \textbf{Left}: We compare the 
  \emph{Planck} and \texttt{P-ACT}
  power spectrum constraints from this work (blue and orange) to the \emph{Planck} $T$- and $E$-mode map-level bounds from \cite{El-Haj:2025zbe} (teal and dark green). Since the matched filter constraints depend on $m_0$, we adopt three characteristic mass values for the localized searches: $m_0=100H_I,200H_I,500H_I$. For lower-mass particles, where the probability of production is enhanced, the power spectrum dominates on nearly all scales. By contrast, for heavier particles (\emph{e.g.}, the dotted pale green curve), the constraints are split across the parameter space, with the matched filter dominating for rare production (large $\eta_*$). While this work explores hotspots produced very late in inflation, down to $\eta_*=1\,\text{Mpc}$, \cite{El-Haj:2025zbe} only constrain hotspots down to $\eta_*=10\,\text{Mpc}$, thus we truncate to $\eta_*\geq 10\,\mathrm{Mpc}$ for this figure. \textbf{Right}: We compare the matched filter and power spectrum approaches for an optimistic CVL experiment up to $\ell_{\rm{max}}=5000$. To place the two methods on even footing, we perform a joint $T+E$ forecast for the matched filter. For extremely massive particles (whose production is rare), the matched filter constraints again dominate, with the power spectrum winning in the limit of numerous particle production (low masses and/or small $\eta_*$).} 
    \label{fig:comparingMFandPowerspec}
\end{figure*}

In the left panel of \autoref{fig:comparingMFandPowerspec}, we directly compare the results of the \emph{Planck} $T$ and $E$-mode searches from \cite{El-Haj:2025zbe} to the power-spectrum-derived \texttt{P-ACT} results. For lower masses, where particle production is numerous, the power spectrum dominates the constraints across nearly all scales. Note, however, that there is some region of the small $\eta_*$ parameter space where the matched filter bounds become stronger. This is likely due to a combination of the noise properties and the relative amplitude of the position-space profile. 

One can also explore the cosmic-variance limit for these two cases. To directly compare methods, we perform a matched filter forecast for a CVL experiment up to $\ell_{\rm{max}}=5000$, including both $T$ and $E$, with the joint significance
\begin{align}
    \label{eq:MMF}
    \left( \sigma^{T\times E} \right)^{-2}=\sum_\ell\sum_{X,Y\in\{T,E\}}\delta X_\ell C_{\ell,\text{Exp}}^{-1,XY}\delta Y_\ell\,,
\end{align}
\noindent where $\delta X_\ell/\delta Y_\ell$ are the CMB profiles in harmonic space, and the sum appearing here replaces the integral in \autoref{eq:MF} as we work in the full-sky limit.  
We compare this prediction to our power spectrum forecast in the right panel of~\autoref{fig:comparingMFandPowerspec}, where we see that for large masses ($m_0 \sim 500 H_I$), the matched filter search dominates on all scales. Conversely, for smaller masses, where particle production is extremely numerous, the power spectrum presents a superior probe.



\section{Summary and Discussion}
\label{sec:summary}

\noindent Non-adiabatic particle production is a feature of many multi-field inflationary models, and exploring its observational impacts in the CMB has garnered significant interest. In this work, we have considered a model in which an extremely massive particle ($m_\chi\sim\mathcal{O}(100 H_I)$) couples to the inflaton field. Due to the large mass of the additional field, particle production is strongly suppressed and occurs only during a brief window of the inflationary epoch where $m_\chi(\phi)$ reaches its minimum. Such production can induce localized features in late-time cosmological observables \cite{Kim:2021ida,kumar2025earlygalaxiesrareinflationary}, which can be constrained through profile-finding techniques, in direct analogy to thermal Sunyaev-Zel'dovich searches \cite{2001ApJ...560L.111H,Munchmeyer:2019kng, Philcox:2024jpd,El-Haj:2025zbe}. While this approach has been shown to place powerful constraints on the scenario, a more conventional power spectrum analysis can potentially yield improved constraints depending on both the features' abundance and amplitude. 

In this work, we have derived the CMB power spectrum arising from the particle-production model and used this to carry out a search for primordial features. We obtained stringent upper limits on the scenario, finding significant gains for small hotspots ($\eta_* \le 30 \, \mathrm{Mpc}$) sourced by the improved power spectrum measurements from ACT DR6. We performed a similar search for these primordial features using the combined \texttt{P-ACT} and \texttt{P-ACT-LB} datasets, finding the tightest upper bounds on the hotspot amplitudes, with a mild hint for non-zero features obtained at $3-10 \, \rm{Mpc}$. Although this tentative preference for a nonzero amplitude is not strong, future experiments will significantly improve the constraining power and could confirm a signal if it is indeed present. For an SO-like experiment, we forecast that the power spectrum method applied in this work will yield $\sim 8\times$ tighter constraints on small-scale hotspots ($\eta_* < 30 \, \mathrm{Mpc}$) than the \texttt{P-ACT} dataset considered herein. If the $\sim 2.5\sigma$ preference for non-zero $g^2P_*$ found in \texttt{P-ACT} is \emph{bona fide}, an SO-like experiment would be able to detect such hotspots at a $3-5 \sigma$ confidence level. Based on the left panel of~\autoref{fig:comparingMFandPowerspec}, however, features at $\eta_* \sim 10 \, \rm{Mpc}$ should already have been detected by the \emph{Planck} matched filter search if the particle mass is sufficiently low ($m_0 \sim 100 H_I$).  This motivates performing a matched filter search on the ACT polarization data, which were not included in \cite{El-Haj:2025zbe}, particularly focusing on small scales $3 \,\, \text{Mpc}\leq\eta_*\leq 10 \,\, \text{Mpc}$. We also recall that a preference for such a feature in the ACT power spectrum likely correlates with the slight mismatch between the preferred spectral index $n_s$ of ACT and \textit{Planck}.  Indeed, inclusion of the feature in the \texttt{P-ACT} fit allows the value of $n_s$ to move back down to its \emph{Planck}-preferred value.


We have additionally compared the results of our search to the constraints attained from a matched filter search using both \emph{Planck} $T$ and $E$-modes \cite{El-Haj:2025zbe}. For lower-mass particles ($m_0\lesssim 200 H_I$), where production is numerous, constraints from the power spectrum dominate by more than an order of magnitude, while for more massive particles the matched filter search performs favorably.

Given the relative simplicity of the inflationary computations for these scenarios (see Appendix~\ref{Appendix:appendix A}) and the separable nature of the $N$-point functions, our constraints could be simply extended to include other low-order correlators, bolstering the power spectrum. 
A natural extension of this work would be to search for periodic particle production in the CMB (\emph{e.g.}, the model of \cite{Munchmeyer:2019kng,Flauger_2017}), whereupon features are generated during multiple bursts of particle production throughout inflation, yielding correlated fluctuations on a range of scales.  We leave these directions for future work.

\begin{acknowledgments}
We are grateful to Soubhik Kumar for useful conversations. HTJ acknowledges support from the Horizon 2020 ERC Starting Grant (Grant agreement No.~849169). Some of the cosmological analyses were performed on the Hawk and Falcon high-performance computing clusters at the Advanced Research Computing at Cardiff (ARCCA). JCH acknowledges support from DOE grant DE-SC0011941, NASA grant 80NSSC23K0463 [ADAP], and the Sloan Foundation. LHA acknowledges support from the Barry Goldwater Scholarship.  This is not an official Simons Observatory collaboration publication.
\end{acknowledgments}

\appendix
\begin{widetext}
\section{\label{Appendix:appendix A} Computing the Particle Production Correlation Functions}

\noindent In the following, we derive a formula for the primordial $N$-point functions induced by the two-particle action (\autoref{eq:2paction}). While for the purposes of the current work we require only the two-point function, it is a useful (and tractable) pedagogical exercise to extend to all $N$. 

To begin, we define the two-particle action which generates \autoref{eq:2paction}, 
    \begin{align}
        S_{2p} &= \int_{\eta_*}^0 \dd \eta\int \dd^3 x \partial_\eta\zeta \bigg[ \frac{m_\chi(\eta)}{H_I}\delta(\vb{x}-\vb{x}_{HS}) +\frac{m_\chi(\eta)}{H_I}\delta(\vb{x}-\vb{x}_{HS}-\mathbf{d} ) \bigg] \\
&=\int_{\eta_*}^0 \dd \eta\int \frac{\dd^3 k}{(2\pi)^3} \partial_\eta\zeta\frac{m_\chi(\eta)e^{i\vb{k}\cdot \vb{x}_{HS}}}{H_I}[1+e^{i\vb{k}\cdot \vb{d}}]\nonumber \,,
    \end{align}
where $\eta_*$ is the conformal time during inflation. This involves the curvature perturbation, $\zeta(\mathbf{x},\eta)$, which can be rewritten in a mode expansion, 
\begin{align}
    \zeta_{\vb{k}}=\frac{H_I^2}{\dot{\phi}_0\sqrt{2k^3}}[(1-ik\eta)e^{ik\eta}a^\dagger_{\vb{k}}+(1+ik\eta)e^{-ik\eta}a_{-\vb{k}}]\,,
\end{align}
assuming a Bunch-Davies initial state. Using the in-in formalism, we can compute a given correlator as a set of ordered time integrals \citep{Maldacena_2003}
    \begin{align}
    & \langle \mathcal{O}(\eta)\rangle = \sum_{N=0}^{\infty}i^N\int_{-\infty}^{\eta} \dd \eta_N\int_{-\infty}^{\eta_N} \dd \eta_{N-1}...\int_{-\infty}^{\eta_2} \dd \eta_1 \times\bra0\big[H_{\rm int}(\eta_1),\big[H_{\rm int}(\eta_2),...\big[H_{\rm int}(\eta_N),\mathcal{O(\eta)}\big]...\big]\big]\ket0\,.
    \label{eq:ininmaster}
\end{align}

To proceed, we note a couple of mathematical facts. First, any commutator obeys the Leibniz rule, 
\begin{align}
    \comm{\widehat{O}}{\widehat{A}_1~\widehat{A}_2~...\widehat{A}_n}=\comm{\widehat{O}}{\widehat{A}_1}\widehat{A}_2~...\widehat{A}_n+\widehat{A}_1\comm{\widehat{O}}{\widehat{A}_2}\widehat{A_3}...\widehat{A_n}+\cdots \widehat{A}_1...\widehat{A}_{n-1}\comm{\widehat{O}}{\widehat{A}_n}\,.
\end{align}
This additionally tells us that if we now consider a sequence of operators $\{\widehat{O}_i\}$, such that $\comm{\widehat{O}_i}{\comm{\widehat{O}_j}{\widehat{A}_k}}=0=\comm{\widehat{A}_i}{\comm{\widehat{O}_j}{\widehat{A}_k}}$ for all $i,j,k$, then a nested commutator will take the form, 
\begin{align}
    \comm{\widehat{O}_1}{\comm{\widehat{O}_2}{...\comm{\widehat{O}_n}{\widehat{A}_1...\widehat{A}_n}}}=\prod_{j=1}^n\sum_{i=1}^n\comm{\widehat{O}_i}{\widehat{A}_j}\,.
\end{align}
This greatly simplifies the structure of our problem. Given the fact that both $\zeta$ and $\partial_\eta\zeta$ are linear in $a_{\mathbf{k}},a^\dagger_{\mathbf{k}}$, we do not need to worry about higher-order commutators. Additionally, we restrict to the fully connected piece of the $N$-point function, which will appear at $N$-th order in \autoref{eq:ininmaster}. 

Inserting our action into the in-in integrand, we find
    \begin{align}
    &\bra0\big[H_{\rm int}(\eta_1),\big[H_{\rm int}(\eta_2),...\big[H_{\rm int}(\eta_N),\mathcal{O}(\eta\rightarrow0)\big]...\big]\big]\ket0\nonumber \\&=\prod_{i=1}^N\int_{\mathbf{p}_i}\frac{m_\chi(\eta_i)e^{i\mathbf{p}_i\cdot\mathbf{x}_{\rm{HS}}}[1+e^{i\mathbf{p}_i\cdot\mathbf{d}}]}{H_I}\bra0\big[\partial_{\eta_1}\zeta_{\mathbf{p}_1}(\eta_1),\big[\partial_{\eta_2}\zeta_{\mathbf{p}_2}(\eta_2),...\big[\partial_{\eta_N}\zeta_{\mathbf{p}_N}(\eta_N),\prod_{j=1}^{N}\zeta_{\mathbf{k}_j}(\eta\rightarrow0)\big]...\big]\big]\ket0 \nonumber \\
&=\prod_{i=1}^N\int_{\mathbf{p}_i}\frac{m_\chi(\eta_i)e^{i\mathbf{p}_i\cdot\mathbf{x}_{\rm{HS}}}[1+e^{i\mathbf{p}_i\cdot\mathbf{d}}]}{H_I} \prod_{j=1}^N\sum_{i=1}^N\comm{\partial_{\eta_i}\zeta_{\mathbf{p}_i}(\eta_i)}{\zeta_{\mathbf{k}_j}} \,.
\label{eq:ininintegrand}
\end{align}
The factorization implies that the full computation involves just a single simple commutator, given by
 \begin{align}
\lim_{\eta\rightarrow0}\comm{\partial_{\eta_1}\zeta_{\mathbf{p}}(\eta_1)}{\zeta_{\mathbf{k}}(\eta)}=-i\left(\frac{H_I^2}{\dot{\phi}_0}\right)^2\delta_{\vb{p}+\vb{k}}\frac{\eta_1}{p}\sin{p\eta_1}\equiv -iQ(\eta_1,p)\delta_{\mathbf{p}+\mathbf{k}}\,.
\end{align}
Next, we note that the last term in \autoref{eq:ininintegrand} is manifestly symmetric in $\eta_i$, which allows us to transform the limits on the time integrals. For any totally symmetric function $K(x_1,...,x_n)$,
\begin{align}
    \int_{x_0}^{x} \dd x_1\int_{x_0}^{x_1}\dd x_2...\int_{x_0}^{x_{n-1}}\dd x_n K(x_1,...,x_n)=\frac{1}{n!}\int_{x_0}^{x}\prod_{i=1}^n\dd x_i K(x_1,...,x_n)\,;
\end{align}
applying to our scenario, we obtain the $N$-point correlator
\begin{align}
\ev{\zeta_{\mathbf{k}_1}\cdots\zeta_{\mathbf{k}_N}}&=\frac{i^N}{N!}\int_{\eta_*}^0\dd \eta_1...\int_{\eta_*}^{0}\dd \eta_N\prod_{i=1}^N\int_{\mathbf{p}_i}\frac{m_\chi(\eta_i)e^{i\mathbf{p}_i\cdot\mathbf{x}_{\rm{HS}}}[1+e^{i\mathbf{p}_i\cdot\mathbf{d}}]}{H_I} \prod_{j=1}^N\sum_{i=1}^N -iQ(\eta_i,p_i)\delta_{\mathbf{p}_i+\mathbf{k}_i}\nonumber  \\
&=\prod_{i=1}^Ne^{-i\mathbf{k}_i\cdot\mathbf{x}_{\rm{HS}}}[1+e^{-i\mathbf{k}_i\cdot \mathbf{d}}]\int_{\eta_*}^{0}\dd \eta_i \frac{m_\chi(\eta_i)Q(\eta_i,\mathbf{k}_i)}{H_I} \nonumber \\
&=\left(\frac{gH_I^2}{\dot{\phi}_0}\right)^N\prod_{i=1}^Ne^{-i\mathbf{k}_i\cdot\mathbf{x}_{\rm{HS}}}[1+e^{-i\mathbf{k}_i\cdot \mathbf{d}}]\left(\frac{\text{Si}(k_i\eta_*)-\sin{k_i\eta_*}}{k_i^3}\right)\,.
\label{eq:allnpoints}
\end{align}
In the first equality we have exploited the symmetry of the momenta, and in the second line we have used the fact that around particle production, $m_\chi(\eta)\simeq\frac{g\dot{\phi}_0}{H_I}\ln{\abs{\eta_*/\eta}}$. The two-point function follows immediately:
    \begin{align}
    \langle \zeta_{\mathbf{k}_1}\zeta_{\mathbf{k}_2}\rangle_{\text{PP}} &=\frac{g^2H_I^4}{\dot{\phi}_0^2k_1^3k_2^3}\big(\text{Si}(k_1\eta_*)-\sin(k_1\eta_*)\big)\big(\text{Si}(k_2\eta_*)-\sin(k_2\eta_*)\big) e^{-i(\mathbf{k_1}+\mathbf{k_2})\cdot\mathbf{x}_{\text{HS}}}\nonumber \\&\times
    [1+e^{-i(\mathbf{k_1}+\mathbf{k_2})\cdot\mathbf{d}}+e^{-i\mathbf{k}_1\cdot\mathbf{d}}+e^{-i\mathbf{k}_2\cdot\mathbf{d}}]\,,
    \label{apcompofpowerspec}
\end{align}
matching \autoref{eq:powerspec}. 

\section{Additional Figures} \label{app:cosmo-figures}

\noindent We present several figures showing the effects of our primordial feature on cosmological parameters. In \autoref{fig:double-triangle-10}, we showed the shifts on cosmological parameters for \texttt{P-ACT} and \texttt{P-ACT-LB} for our feature at $\eta_* = 10 \, \rm{Mpc}$; below, we show the same figure for features at $1$, $100$, and $1000 \, \rm{Mpc}$. We note that for features larger than the pivot scale $k_0 = 0.05 \, \rm{Mpc}^{-1}$ ($\eta_* > 20 \, \rm{Mpc}$), the inclusion of the feature does not cause a major shift in the main cosmological parameters. 
Although not of interest for our main search, we additionally include a figure that shows the effect of including our primordial hotspot feature on the quantities $H_0$, $\Omega_m$, and $S_8 \equiv \sigma_8 \sqrt{\Omega_m/{0.3}}$ in \autoref{fig:tensions}.

\begin{figure*}
    \centering
    \includegraphics[width=\linewidth]{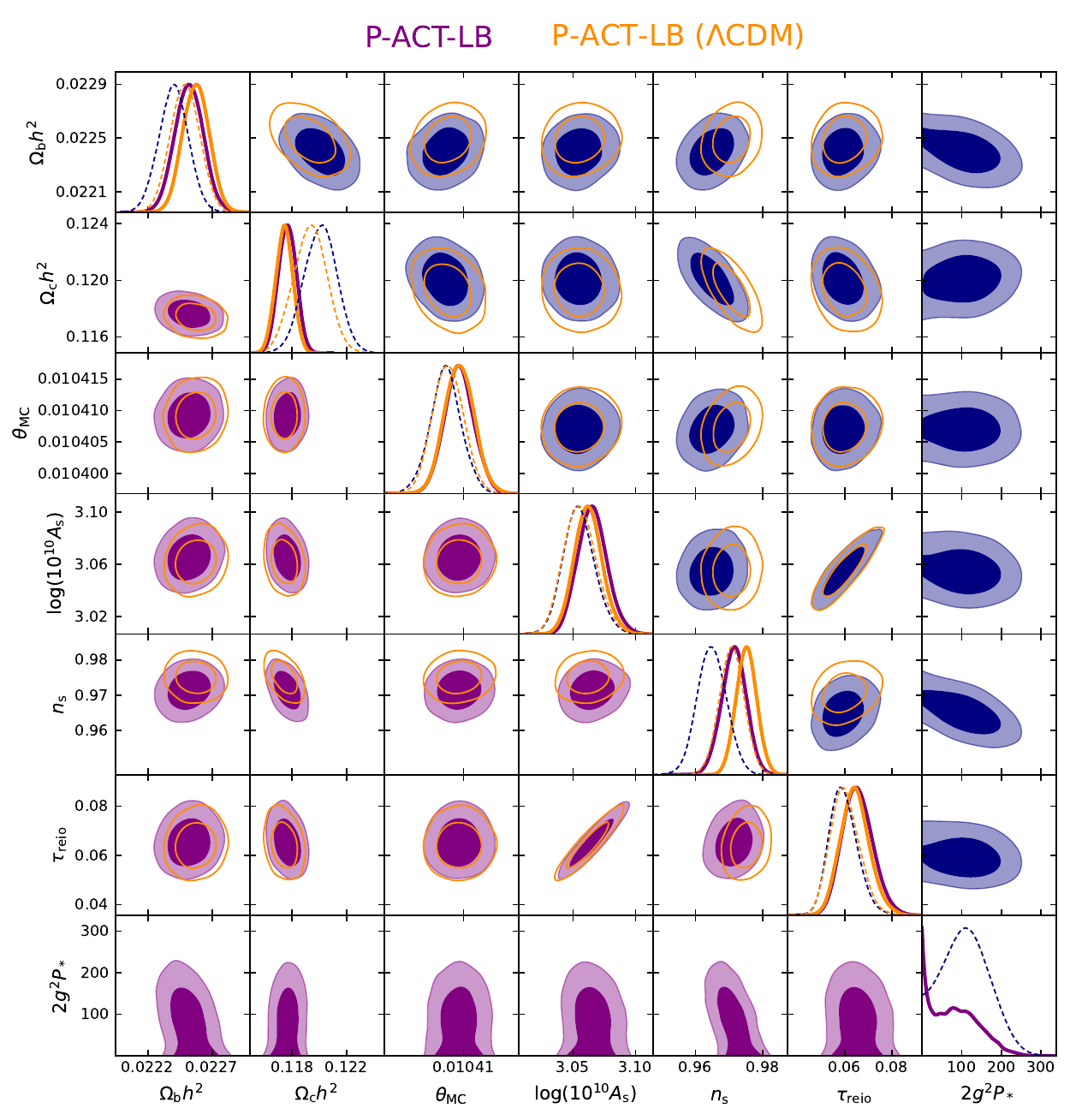}
    \caption{As in~\autoref{fig:double-triangle-10}, but for $\eta_* = 1 \, \rm{Mpc}$. Note that the hotspot feature lies on the boundary of the observable window, and is thus very poorly constrained.}
    \label{fig:double-triangle-1}
\end{figure*}

\begin{figure*}
    \centering
    \includegraphics[width=\linewidth]{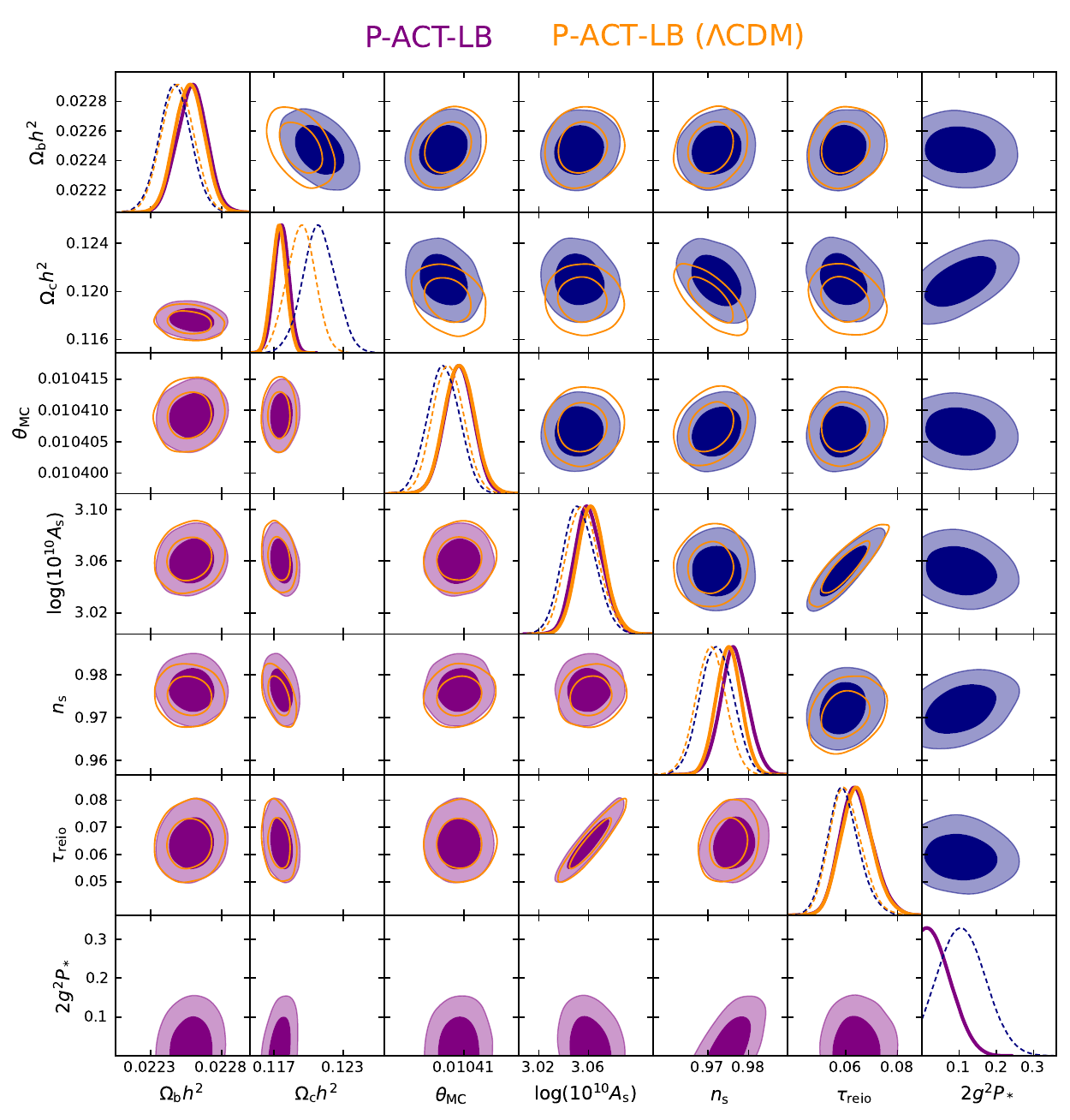}
    \caption{As in~\autoref{fig:double-triangle-10}, but for $\eta_* = 100 \, \rm{Mpc}$.}
    \label{fig:double-triangle-100}
\end{figure*}

\begin{figure*}
    \centering
    \includegraphics[width=\linewidth]{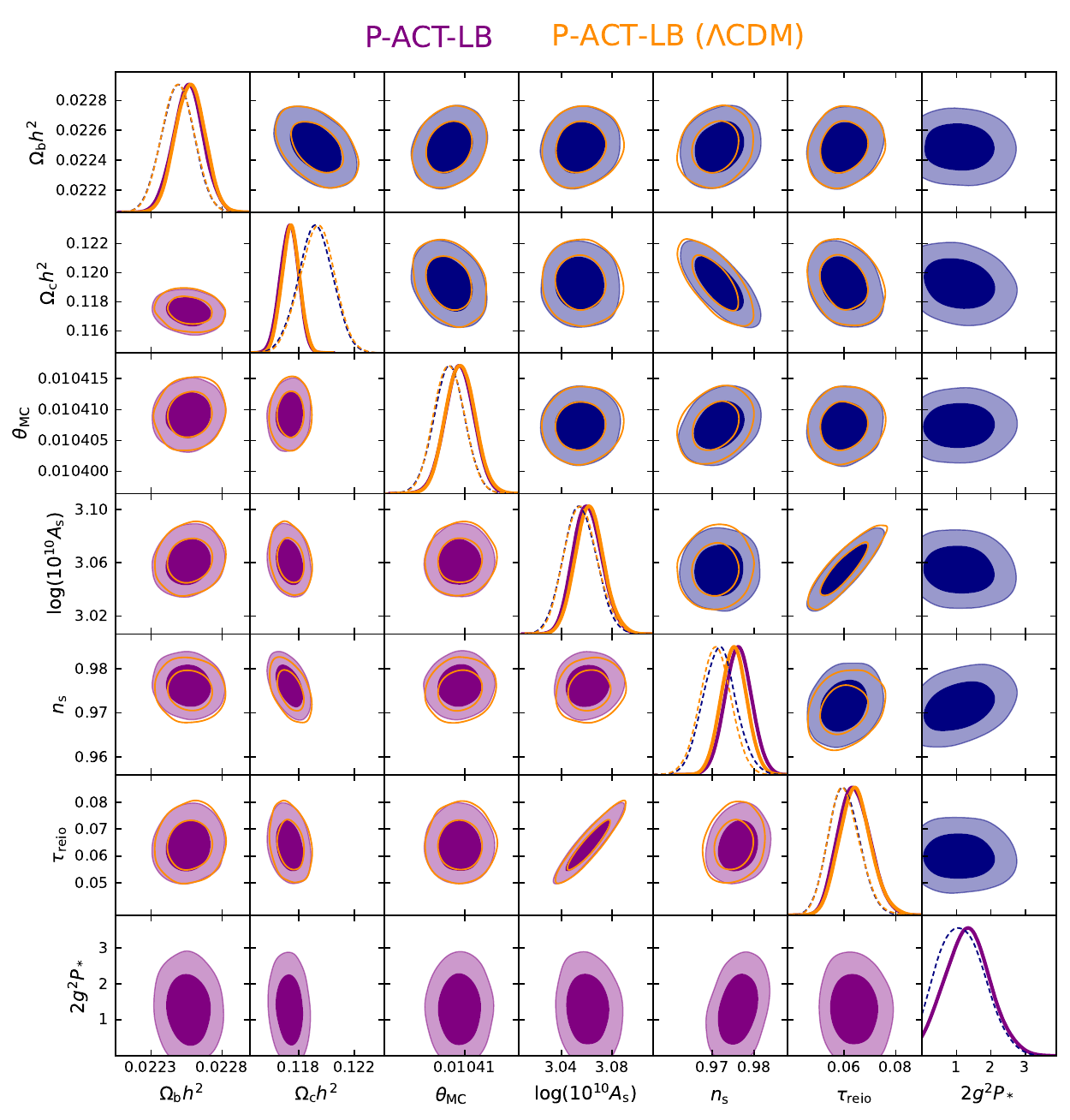}
    \caption{As in~\autoref{fig:double-triangle-10}, but for $\eta_* = 1000 \, \rm{Mpc}$.}
    \label{fig:double-triangle-1000}
\end{figure*}

\end{widetext}

\begin{figure}
    \centering
    \includegraphics[width=\linewidth]{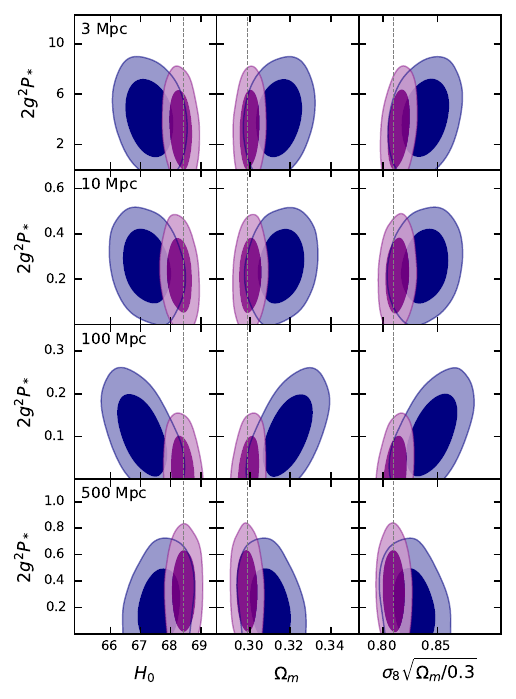}
    \caption{The effects of our primordial hotspot feature on $H_0$, $\Omega_m$, and $S_8$. The scales chosen represent the range of observable scales for our model.  The blue (purple) contours show results for \texttt{P-ACT} (\texttt{P-ACT-LB}).  The vertical dashed line represents the best-fit value for $\Lambda$CDM using \texttt{P-ACT-LB}. No significant shifts are seen with respect to the $\Lambda$CDM inference of these parameters. }
    \label{fig:tensions}
\end{figure}

\vspace{24pt}

\bibliography{Hotspot_paper}

\end{document}